\documentclass[sigconf]{acmart}

\AtBeginDocument{%
  }

\setcopyright{acmlicensed}
\copyrightyear{2018}
\acmYear{2018}
\acmDOI{XXXXXXX.XXXXXXX}
\acmConference[Conference acronym 'XX]{Make sure to enter the correct
  conference title from your rights confirmation email}{June 03--05,
  2018}{Woodstock, NY}
\acmISBN{978-1-4503-XXXX-X/2018/06}
\newtheorem{definition}{Definition}[section]
\newtheorem{assumption}{Assumption}[section]
\usepackage{graphicx}
\usepackage{caption}
\usepackage{subcaption} 
\graphicspath{{figs/}}   
\usepackage{graphicx}
\usepackage{float} 
\usepackage{stfloats}
\usepackage{amsthm}
\usepackage{booktabs}
\usepackage{multirow}
\usepackage{tabularx}
\usepackage{makecell}
\usepackage{siunitx}   
\usepackage{pgfplots}  
\usepackage{array}
\usepackage{caption}
\usepackage{hyperref}

\usepackage{amsmath, amssymb}
\usepackage{graphicx}
\usepackage{adjustbox}
\pgfplotsset{compat=1.18}
\usepackage{xcolor}
\usepackage{enumitem}
\usepackage{threeparttable} 
\usepackage{amsmath,amssymb}
\usepackage{algorithm}
\usepackage{algorithmicx}
\usepackage{algpseudocode}
\usepackage{tablefootnote}
\usepackage{xcolor}
\usepackage[most]{tcolorbox} 
\usepackage{microtype}       
\usepackage{xurl}            
\usepackage{seqsplit}        

\usepackage{booktabs}
\usepackage{multirow}
\usepackage{adjustbox}
\usepackage{threeparttable} 
\usepackage{array}
\usepackage{url}            
\newcommand{\sd}[1]{{\small $\pm$ #1}}

\tcbset{
  colback=gray!5!white,
  colframe=gray!40!black,
  boxrule=0.3pt,
  arc=0.8mm,
  left=1mm,right=1mm,top=1mm,bottom=1mm,
  width=\columnwidth, 
  boxsep=1pt,
  enhanced,
  breakable,
  fonttitle=\bfseries,
  coltitle=black,
  before skip=6pt,
  after skip=6pt
}

\floatname{algorithm}{Algorithm}

\makeatletter
\@ifundefined{Bbbk}{}{}
\makeatother
\usepackage{amsmath}

\usepackage{seqsplit} 
\newcolumntype{T}{>{\ttfamily\small\raggedright\arraybackslash}X} 
\newcolumntype{L}{>{\raggedright\arraybackslash}X}                

\newcolumntype{Y}{>{\raggedright\arraybackslash}X}

\newcolumntype{Y}{>{\raggedright\arraybackslash}X}

\begin{document}

\title[Towards Compositional Generalization in LLMs for Smart Contract Security]{Towards Compositional Generalization in LLMs for Smart Contract Security: A Case Study on Reentrancy Vulnerabilities}

\author{Ying Zhou\textsuperscript{\textdagger}, Jiacheng Wei\textsuperscript{\textdagger}, Yu Qi, Faguo Wu\textsuperscript{*}, Xiao Zhang\textsuperscript{*}}
\thanks{ 
    \textsuperscript{\textdagger} Equal contribution. \\
    \textsuperscript{*} Corresponding authors.
}
\affiliation{
    \institution{School of Artificial Intelligence, Beijing, China}
    \city{}
    \country{}
}
\affiliation{
    \institution{Beijing Advanced Innovation Center for Future Blockchain and Privacy Computing, Beijing, China}
    \city{}
    \country{}
}
\email{zy2442115@buaa.edu.cn; jakiewei258@gmail.com}

\renewcommand{\shortauthors}{Trovato et al.}

\begin{abstract}
Large language models (LLMs) demonstrate remarkable capabilities in natural language understanding and generation. Despite being trained on large-scale, high-quality data, LLMs still fail to outperform traditional static analysis tools in specialized domains like smart contract vulnerability detection. To address this issue, this paper proposes a post-training algorithm based on atomic task decomposition and fusion. This algorithm aims to achieve combinatorial generalization under limited data by decomposing complex reasoning tasks. Specifically, we decompose the reentrancy vulnerability detection task into four linearly independent atomic tasks: identifying external calls, identifying state updates, identifying data dependencies between external calls and state updates, and determining their data flow order. These tasks form the core components of our approach. By training on synthetic datasets, we generate three compiler-verified datasets. We then employ the Slither tool to extract structural information from the control flow graph and data flow graph, which is used to fine-tune the LLM's adapter. Experimental results demonstrate that low-rank normalization fusion with the LoRA adapter improves the LLM's reentrancy vulnerability detection accuracy to 98.2\%, surpassing state-of-the-art methods. On 31 real-world contracts, the algorithm achieves a 20\% higher recall than traditional analysis tools. 
\end{abstract}

\ccsdesc[500]{Do Not Use This Code~Generate the Correct Terms for Your Paper}
\ccsdesc[300]{Do Not Use This Code~Generate the Correct Terms for Your Paper}
\ccsdesc{Do Not Use This Code~Generate the Correct Terms for Your Paper}
\ccsdesc[100]{Do Not Use This Code~Generate the Correct Terms for Your Paper}

\keywords{large language models, compositional generalization, smart contract security, reentrancy vulnerabilities}

\received{20 February 2007}
\received[revised]{12 March 2009}
\received[accepted]{5 June 2009}

\maketitle
\section{Introduction}
Large language models (LLMs), pre-trained on extensive text corpora, exhibit remarkable capabilities in both language understanding and generation. This arises because abundant training data expand the parameter space far beyond the constraining signal. At the same time, the deterministic rank of the data distribution provides structural regularity, guiding the model toward stable solutions. Together, these conditions enable emergent capabilities. However, in specialized vertical domains such as smart contract vulnerability detection, even advanced models like GPT-5 fail to outperform static analysis tools~\cite{ince2025generative, sun2024gptscan}. This remains the case despite GPT-5 being trained on nearly exhaustive high-quality data. Therefore, data from many vertical domains are insufficient to support emergent behavior in LLMs, let alone complex reasoning tasks requiring high accuracy. Furthermore, LLMs constrained by frozen pre-training, often lag behind the rapidly evolving data of vertical domains~\cite{veldanda2024llm, nishu2025dense}. Tasks such as code verification are inherently constrained by graph structures, which encode control-flow and data-flow dependencies~\cite{yamaguchi2014modeling, wi2022hiddencpg} and ultimately determine execution correctness.

Combinatorial generalization refers to a model’s ability to transfer capabilities across changing data distributions (i.e., new and unseen combinatorial structures) by leveraging its understanding of the underlying components. In recent years, extensive research~\cite{anonymous2025combination, yang2025heuragenix} has investigated the combinatorial generalization of LLMs, particularly in the vertical context of complex reasoning. Many studies~\cite{xin2025atomr, teng2025atom, liu2025chaos, li2025reasoning} have sought to reduce the complexity of reasoning tasks by decomposing them into simpler subtasks to improve accuracy. This decomposition has further motivated research on how to recombine these subtasks into the overall task. However, current research~\cite{christofidellis2023unifying, li2019compositional, tian2023decompose, jothimurugan2023robust} has not adequately addressed the stability and efficiency of models when applied to novel tasks. Furthermore, little research has examined the combinatorial generalization of LLMs in specialized vertical domains, such as code vulnerability detection. Smart contract vulnerabilities are closely tied to asset security and have thus attracted more research attention and tool development than traditional software vulnerabilities. Among them, reentrancy vulnerabilities~\cite{cecchetti2021compositional}, as one of the most representative and persistent types, are particularly well-suited for studying combinatorial generalization. Detecting such vulnerabilities requires integrating multiple factors, including external calls, state updates, execution order, data dependencies, and the absence of defense mechanisms. Therefore, this article investigates \textbf{how atomic task decomposition and recombination can enhance LLMs in detecting reentrancy vulnerabilities under limited data}, thereby advancing out-of-distribution generalization.

A reentrancy vulnerability occurs when a smart contract relies on a previously read state during an external call, and the call precedes the state update in the data flow. Based on this definition, we decompose reentrancy vulnerabilities into four linearly independent atomic tasks: (1) identifying external calls, (2) identifying state updates, (3) detecting data dependencies between external calls and state updates, and (4) determining their data-flow order. Preliminary evaluation shows that LLMs perform poorly except (2) identifying state updates. To overcome dataset limitations, we used LLMs to generate prompts guided by carefully designed rules. With this approach, we constructed compilable datasets for the three tasks. We then employed Slither to compile these datasets, extracting knowledge from the intermediate representation (IR) layer and generating structural information for the Control Flow Graph (CFG) and Data Flow Graph (DFG) subtasks. This process ultimately yielded three new CoT datasets. We subsequently fine-tuned the base model to develop task-specific adapters. To improve LLMs accuracy in reentrancy vulnerability detection, we fused the three adapters and the base-model (represent the state update factor). The fusion process was optimized with cross-entropy loss and accuracy as objective functions, yielding a final adapter specialized for reentrancy vulnerabilities.


In our experiments, LoRA adapters trained on compiler-verified factor datasets achieve near-saturated results (97–99\% F1). When fused, the adaptive variant reaches 94.7\% F1 and 98.2\% ACC, improving by 16.8\% and 5.7\% over the single-task LoRA baseline. On 31 real contracts, recall increases to 87.1\%, exceeding the best traditional analyzer by 23.77\%. Full-rankness tests under class-prior perturbations show nearly identical AUROC and AUPRC curves, confirming balanced and stable factor fusion. These results demonstrate that compositional fusion effectively unifies factor-level reasoning and achieves robust generalization beyond fine-tuned and rule-based baselines.


In summary, the contributions of this paper are shown as follows:
\begin{itemize}
    \item For complex reasoning tasks in vertical domains, we propose a post-training algorithm that leverages atomic task decomposition and fusion to realize compositional out-of-distribution generalization, allowing LLMs to generalize under limited and sparse data conditions.
    \item As a concrete instantiation, we apply the proposed algorithm to reentrancy vulnerability detection by decomposing the task, and fine-tuning on them, thereby validating its effectiveness.
    \item We build three compiler-verified datasets (external call task, dependency task, order task, about 2.5k cases each) with CFG/DFG cues and semantic refinement, plus 31 real vulnerable contracts for realistic end-to-end evaluation.
    \item Our method achieves 98.2\% reentrancy vulnerability detection accuracy, surpassing state-of-the-art performance. Applied to real-world vulnerable contracts, it outperforms traditional static analysis tools, exceeding Slither's highest accuracy by 20\%.
\end{itemize}

\section{Preliminary}
\subsection{Out-of-Distribution Generalization}
Out-of-distribution generalization means that a trained model can correctly answer questions about combinations not included in the training set. In simple terms, this requires the model to learn the underlying logic of the data rather than just memorize it. Following the idea of curriculum learning, which breaks knowledge into small units and learns it step by step with increasing difficulty, we assume that \textit{any reasoning task $\mathcal{F} (r)$ can be decomposed into atomic and independent units $a_{i}$}.The formula is as follows,
\begin{equation}
    \mathcal{F} (r)=\prod_{i=1}^{n} a_{i}, \quad a_{i}\bot a_{j}(i\ne j)
\label{decompose}
\end{equation}

We assume that the size of the training dataset is positively correlated with task complexity, that is, $\left| DS \right| \propto \mathrm{C}(r)$. As shown in Equation~\ref{decompose}, if a reasoning task $r$ with complexity $\mathcal{C}(r)=N $ is decomposed into $n$ independent atomic subtasks, the average complexity of each subtask becomes $\sqrt[n]{N} $, and the total complexity is $n\sqrt[n]{N}$. Therefore, the size of the training dataset is reduced from $\mathcal{O}(N)$ to $\mathcal{O}(n\sqrt[n]{N})$. This alleviates the problem of dataset scarcity.

Learning from the training results of atomic tasks can effectively enhance the learning of the overall reasoning task. Specifically, the weight parameters $\theta_i$, obtained by training the model $f_i$ on dataset $D_i$ for subtask $i$, minimize the average loss $\mathcal{L}$ between the predictions and the true labels. The formula is given as follows,

\begin{equation}
\theta_i\;=\; \arg\min_{\theta_i} \;
\mathbb{E}_{(x,y)\sim D_i}\!\left[ \, \mathcal{L} \left(f_i(x;\theta_i),\,y\right) \, \right]
\label{train}
\end{equation}

As shown in Equation~\ref{train}, the training results of atomic reasoning need to be integrated into the weight parameters of the overall reasoning task in order to achieve high accuracy with simple training on a small dataset. Specifically, a non-negative and normalized weight distribution $\alpha_i(x)$ is maintained. The output probabilities of multiple subtasks are weighted and combined to obtain the prediction $\hat{p}(y \mid x)$ for the overall task. The formula is:

\begin{equation}
\hat{P}(y \mid x) \;=\; \sum_{i=1}^n \alpha_i(x)\, p_i(y \mid x),
\; \sum_{i=1}^n \alpha_i(x) = 1,\; \alpha_i(x)  \ge 0.
\label{fusion}
\end{equation}

\subsection{Reentrancy vulnerability}
\begin{figure}[t]
  \centering
  \includegraphics[width=\linewidth]{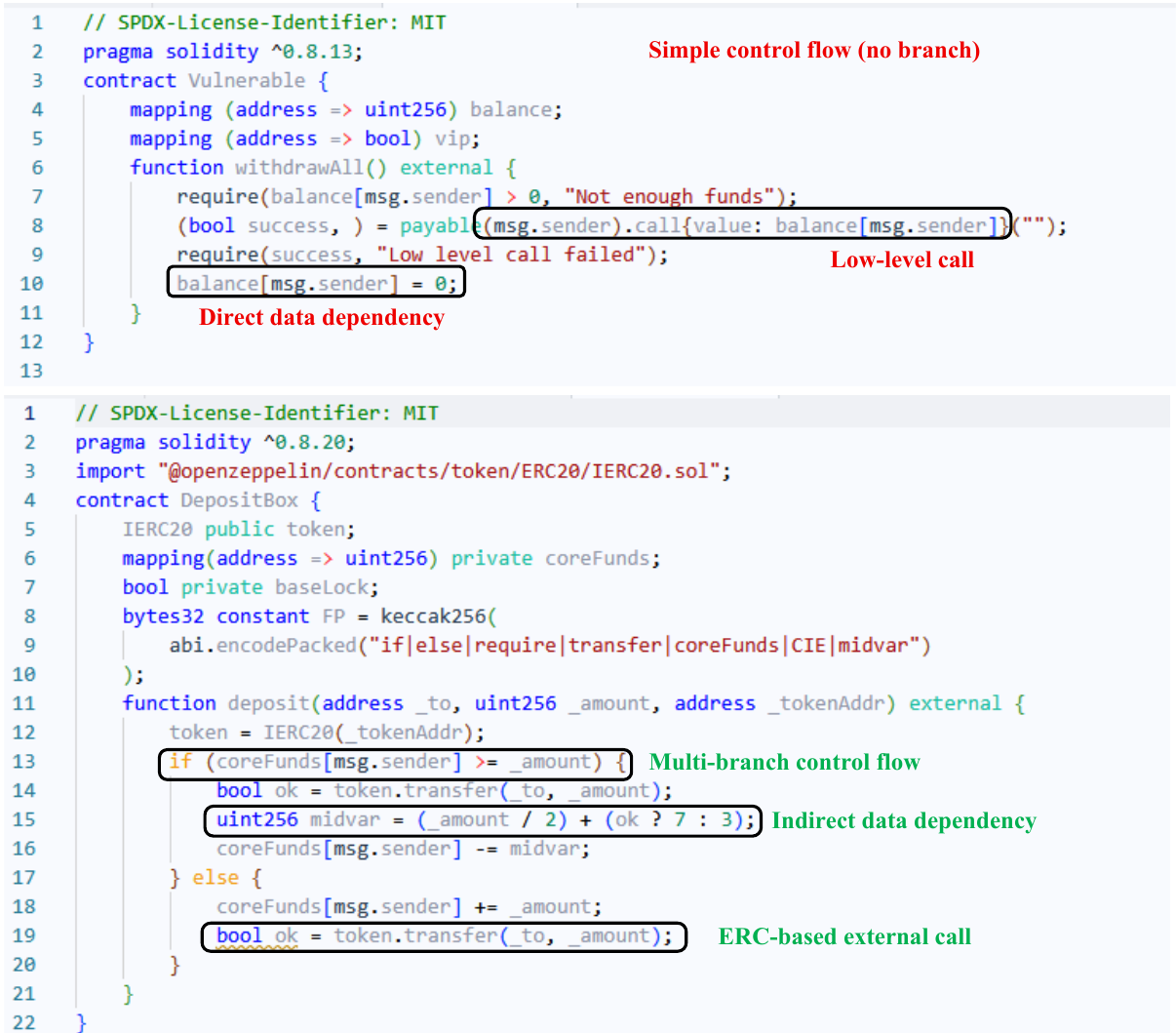}
  \caption{
  Comparison of a typical reentrancy pattern (top) and our dataset-constructed example (bottom).
  }
  \label{fig:code_example}
  \vspace{-1em}
\end{figure}

Code vulnerability detection places greater demands on LLMs than code generation, as it requires a deep understanding of program logic. For smart contracts, whose correctness directly affects financial security, such reasoning is crucial. Reentrancy vulnerabilities, among the earliest and most well-known flaws, have caused millions of dollars in losses (e.g., the DAO attack~\cite{feichtinger2024sok}). Their root cause lies in external calls that transfer control without timely state updates, allowing attackers to reenter and drain assets. Yet, existing detection tools often depend on surface-level syntactic heuristics, such as recognizing \texttt{.call()} followed by a balance update. In reality, many recent vulnerabilities stem from misuse of ERC-standard APIs (e.g., \texttt{safeTransferFrom}) rather than legacy \texttt{call.value()}. Figure~\ref{fig:code_example} illustrates this contrast, showing how our dataset includes ERC-based interactions and multi-branch control flows for higher structural complexity. Therefore, given a compilable smart contract program $P$ with $N$ lines, the reentrancy vulnerability detection task can be decomposed into four subtasks.

\begin{definition}[External-call factor.]
The external call(e.g., \texttt{call}, \texttt{delegatecall}, or ERC-interface call) factor is,
\begin{equation}
   \phi_E : \mathcal{P} \to \{0,1\}^N, \phi_E(P)[i] = 1 \iff \text{line } i \text{ is external call.}
\end{equation}
\label{ec}
\end{definition}

\begin{definition}[State-update factor.]
The state update (e.g., ) factor is defined as,
\begin{equation}
   \phi_S : \mathcal{P} \to \{0,1\}^N, \phi_S(P)[i] = 1 \iff \text{line } i \text{ is state update.}
\end{equation}
\label{su}
\end{definition}

\begin{definition}[Dependency factor.]
If the state update at line $i$ writes variables that are read by the external call at line $j$, a data dependency occurs and can be formulated as:
\begin{equation}
\begin{aligned}
    \phi_D : \mathcal{P} \to \{0,1\}^{N\times N},\phi_D(P)[i,j] = 1 \iff \\ \phi_S(P)[i]=1 \wedge \phi_E(P)[j]=1 \wedge \text{vars}(i)\cap  \text{vars}(j) & \neq  0.
\end{aligned}
\end{equation}
\label{D}
\end{definition}

\begin{definition}[Ordering factor.]
The relative order between state updates and external calls in the data flow is defined as the order factor:
\begin{equation}
\begin{aligned}
\phi_O &: \mathcal{P} \to \{-1,0,+1\}^{N\times N}, \\[3pt]
\phi_O(P)[i,j] &=
\begin{cases}
+1, & \text{if } \phi_D(P)[i,j] = 1 \ \wedge\ i \prec_{\mathrm{df}} j, \\[3pt]
-1, & \text{if } \phi_D(P)[i,j] = 1 \ \wedge\ j \prec_{\mathrm{df}} i, \\[3pt]
0,  & \text{otherwise.}
\end{cases}
\end{aligned}
\end{equation}
where $i \prec_{\mathrm{df}} j$ indicates that there exists a feasible path in the data flow graph from line $i$ to line $j$  without being overwritten.
\label{O}
\end{definition}

\section{Methodology}

\begin{figure}[t]
  \centering
  \includegraphics[width=\linewidth]{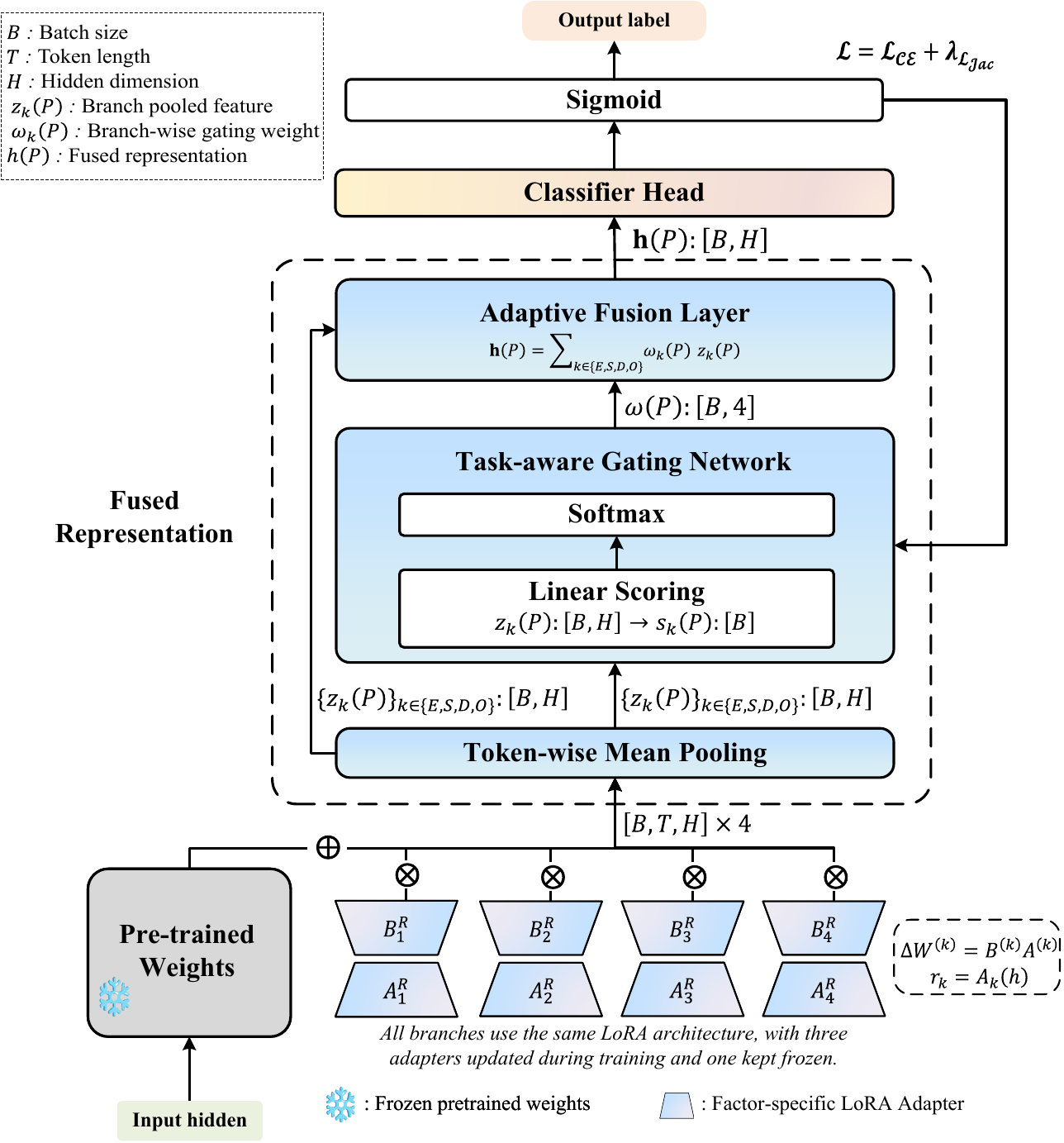}
  \caption{Overall framework of CompFuse.}
  \label{fig:framework}
\end{figure}

\subsection{Compositional Function Modeling}
\label{subsec:cei-function}
For program $P$, a reentrancy vulnerability exists only when the following four conditions are satisfied,
\begin{equation}
\label{eq:cei-boolean}
\exists (i, j) \in \mathcal{N}\!\times\!\mathcal{N} \quad 
\text{s.t.} \;\left\{
\begin{aligned} 
&\phi_E(P)[i] = 1\\
&\phi_S(P)[j] = 1\\ 
&\phi_D(P)[i,j] = 1\\
&\phi_O(P)[i,j] = -1
\end{aligned}
\right.
\Rightarrow
\text{ReVul}(P)=\text{True}  \\
\end{equation}

We map the discrete order factor $\phi_O(P)[i,j] \in {-1, 0, +1}$ to a continuous risk weight using a sigmoid relaxation, formulated as:

\begin{equation}
\label{risk}
\tilde{\phi}_O(P)[i,j]  = \frac{1}{1 + \exp\!\big(\alpha\,\phi_O(P)[i,j]\big)}, \; \alpha > 0
\end{equation}

where $\alpha>0$ controls the mapping sharpness, and $\tilde{\phi}_O(P)[i,j]\in[0,1]$ represents the continuous risk weight derived from the discrete order factor $\phi_O(P)[i,j]$. 

We replace the discrete criterion in Eq.~\ref{eq:cei-boolean} with a temperature-controlled differentiable approximation to achieve stable and end-to-end differentiable optimization with non-vanishing gradients. Also, using the continuous risk weight of Eq.~\ref{risk} to obtain the formula,

\begin{equation}
\label{soft}
\begin{aligned}
f(P) = \log \sum_{i,j}
\exp\Big(
\phi_E(P)[i]\cdot
\phi_S(P)[j]\cdot
\phi_D(P)[i,j]\cdot
\tilde{\phi}_O(P)[i,j]
\Big)
\end{aligned}
\end{equation}

We apply a sigmoid transformation to convert the differentiable score $f(P)$ in Eq.~\ref{soft} into a binary prediction for discrete interpretability during inference. Here, $\sigma$ denote the sigmoid and $\text{softmax}_{\tau }$ denote the softmax with temperature $\tau$. The formula is,

\begin{equation}
\label{eq:pred_def}
\hat{y} = \sigma\!\big(\alpha f(P) - \tau\big)
\end{equation}

where $\sigma$ denotes the sigmoid function that converts the continuous score $f(P)$ into a probabilistic prediction $\hat{y}\in[0,1]$.

In summary, the overall modeling process originates from the functional decomposition of the reentrancy definition. By ensuring sufficient coverage of sub-task combinations in the training set, the Jacobian of the composite function $f(P)$ becomes full-rank across all directions, which enables compositional generalization. This property demonstrates the theoretical soundness and completeness of the proposed design. A detailed analysis of the dataset requirements for each subtask is provided in Subsection~\ref{analysis}.

\subsection{Combination Fusion Modeling and Theoretical Feasibility}
\label{subsec:fusion}

Given an input program $P$, the frozen backbone with factor-specific adapters produces last-layer hidden states $H_k(P)\in\mathbb{R}^{T\times H}$ for four branches $k\in\{E,S,D,O\}$, where $B$, $T$, and $H$ denote the batch size, token length, and hidden dimension. Token-wise Mean Pooling yields per-branch features,
\begin{equation}
\label{eq:z}
z_k(P)=\mathrm{MeanPool}\!\big(H_k(P)\big)\in\mathbb{R}^{H}.
\end{equation}

The Task-aware Gating Network assigns normalized weights via a temperature-controlled Softmax. Given branch scores $s_k(P)$ and temperature $\tau>0$,
\begin{equation}
\label{eq:omega}
\omega_k(P)=
\frac{\exp\!\big(s_k(P)/\tau\big)}
{\sum_{j\in\{E,S,D,O\}}\exp\!\big(s_j(P)/\tau\big)},
\;
\sum_{k}\omega_k(P)=1,\ \omega_k(P)\ge0.
\end{equation}

when $\tau\!\to\!0$, $\alpha_k$ becomes one-hot on $\arg\max_j s_j(P)$, while a large $\tau$ approaches uniform mixing over active branches. 

The Adaptive Fusion Layer then forms the fused representation as a convex combination of factor features.

\begin{equation}
\label{eq:h}
\mathbf{h}(P)=
\sum_{k\in\{E,S,D,O\}}\omega_k(P)\,z_k(P)\in\mathbb{R}^{H}.
\end{equation}

The classifier head implements the scoring function $f(P)$ using a linear projection over the fused representation.
The prediction probability $\hat{y}$ is obtained from $f(P)$ following Eq.~\eqref{eq:pred_def}.

\begin{equation}
\label{eq:pred}
f(P)=\langle w,\,\mathbf{h}(P)\rangle+b.
\end{equation}

To align routing with functional relevance, a Jacobian-based objective is introduced. Let $f_y(P)$ be the logit of the ground-truth class. The per-branch sensitivity and its normalized target are
\begin{equation}
\label{eq:jac}
s^{J}_k(P)=\big\|\nabla_{z_k} f(P)\big\|_2,
\qquad
\tilde{\pi}_k(P)=\frac{s^{J}_k(P)}{\sum_j s^{J}_j(P)}.
\end{equation}

The Jacobian alignment minimizes the divergence between the gating distribution and the sensitivity target,
\begin{equation}
\label{eq:jac_align}
\mathcal{L}_{J}=
\mathrm{KL}\!\big(\tilde{\pi}(P)\,\|\,\omega(P)\big).
\end{equation}

The overall objective uses only cross-entropy and Jacobian alignment,
\begin{equation}
\label{eq:loss}
\mathcal{L}=\mathbb{E}\Big[\mathrm{CE}\big(y,\hat{y}(P)\big)+\lambda_J\,\mathcal{L}_{J}\Big],
\end{equation}
where $\lambda_J\ge 0$ controls the strength of alignment. All factor adapters remain frozen; only the gating parameters and the classifier head are optimized.

\subsection{Analysis of our Methods based on  Out-of-Distribution}
\label{analysis}
Building on theoretical insights from~\cite{zhang2025complexity}, models with constrained complexity tend to learn reasoning-based compositional rules rather than merely memorizing superficial input–output mappings. Furthermore, the analysis of Transformer learning dynamics suggests that sufficient compositional coverage in the training distribution is crucial. Under Lipschitz or norm-based regularization, the model’s hypothesis space tends to capture a structured reasoning function that generalizes to unseen compositions. Therefore, we argue that to enable out-of-distribution compositional generalization, the synthetic dataset should satisfy the following four sufficient conditions:

\begin{definition}[Smoothness]
    $f(P)$ and $\phi$ are all Lipschitz continuous,
    \begin{equation}
        \begin{aligned}
             \|f(P_1) - f(P_2)\| \le L_f \|P_1 - P_2\|,\\
             \|\phi_{E,S,D,O}(P_1) - \phi_{E,S,D,O}(P_2)\| \le  L_\phi &\|P_1 - P_2\|
        \end{aligned}
    \end{equation}
    where $L_f$ and $L_\phi$ denote the Lipschitz constant and .
\end{definition}

\begin{definition}[Compositionality]
    The test distribution $Q$ lies within the compositional closure $f(P)$ of the training distribution $P$, 
    \begin{equation}
        \mathrm{supp}(Q) \subseteq f(P)
    \end{equation}
    where $\mathrm{supp}(Q)$ denotes the support set of test distribution $Q$.
\end{definition}

\begin{definition}[Full-rankness]\label{def:rankness}
    The Jacobian of the ground-truth function $f(P)$ with respect to latent factors $z$ is full rank, 
    \begin{equation}
            \mathrm{rank}\!\left(\nabla_z f(P)\right) = d_z, \quad \forall x \in \mathrm{supp}(P)
    \end{equation}
    where $d_z$ denotes the dimensionality of the latent space.
\end{definition}

\begin{definition}[Consistency]
    There exists a reference point $p^0 \in \mathrm{supp}(P)$ ensuring integral consistency:
    \begin{equation}
        \begin{aligned}
            f(P) = f(p^0) + \int_{p^0}^{P} \nabla f(u)\,du, \\
            \phi_{E,S,D,O}(P) = \phi_{E,S,D,O}(p^0) +  \int_{p^0}^{P} \nabla & \phi_{E,S,D,O}(u)\,du
        \end{aligned}
    \end{equation}
    This enforces continuity of the learned mapping over the support domain and guarantees a smooth functional extension from training to test regions.
\end{definition}

The modeling design, fusion algorithm, synthetic dataset construction, and the overall training pipeline are all designed to satisfy the four conditions outlined above.

\begin{theorem}[Smoothness Satisfaction]
Under the modeling design of Eq.~\ref{soft}, the smoothness condition is guaranteed.
\end{theorem}

See Appendix B.1 for details on the proof process.

\begin{theorem}[Compositionality by Design]
Under Assumptions~\ref{asu1} and~\ref{asu2}, the compositionality condition  is guaranteed.
\end{theorem}

See Appendix B.2 for details on the proof process.

\begin{theorem}[Full-rankness Satisfaction]
Under the construction of the synthetic dataset and the proposed fusion architecture, the full-rankness condition is satisfied.
\end{theorem}

See Appendix B.3 for details on the proof process.

\begin{theorem}[Consistency Satisfaction]
Under the proposed modeling and fusion design, the consistency condition is satisfied.
\end{theorem}

See Appendix B.4 for details on the proof process.

\section{Dataset}
To address annotation scarcity, we adopt a hybrid paradigm integrating seed labels, controlled synthesis, and preference refinement, inspired by instruction-tuning frameworks~\cite{weifinetuned}.
Existing vulnerability datasets for smart contracts are mostly label-level, offering only binary annotations of vulnerable or non-vulnerable cases.  Such supervision lacks the semantic granularity needed for compositional reasoning, leaving models unable to capture how external calls, state updates, and execution order jointly determine reentrancy behavior.  Moreover, no dedicated factor-level datasets exist, and constructing them manually is prohibitively expensive.  To enable future compositional modeling, a scalable and systematic approach for generating such data is essential.

\subsection{External Call Dataset}
The first dataset instantiates the hybrid paradigm for the task of \emph{external call identification}, a critical step in modeling reentrancy vulnerabilities. Following the principles in Section~\ref{analysis}, the construction proceeds in three stages—expert seeds ensure correctness, controlled synthesis expands ERC-based coverage, and refinement enhances semantic reliability—forming the foundation for subsequent dependency and reentrancy tasks.

\begin{figure*}[!t]
  \centering
  \includegraphics[width=\textwidth]{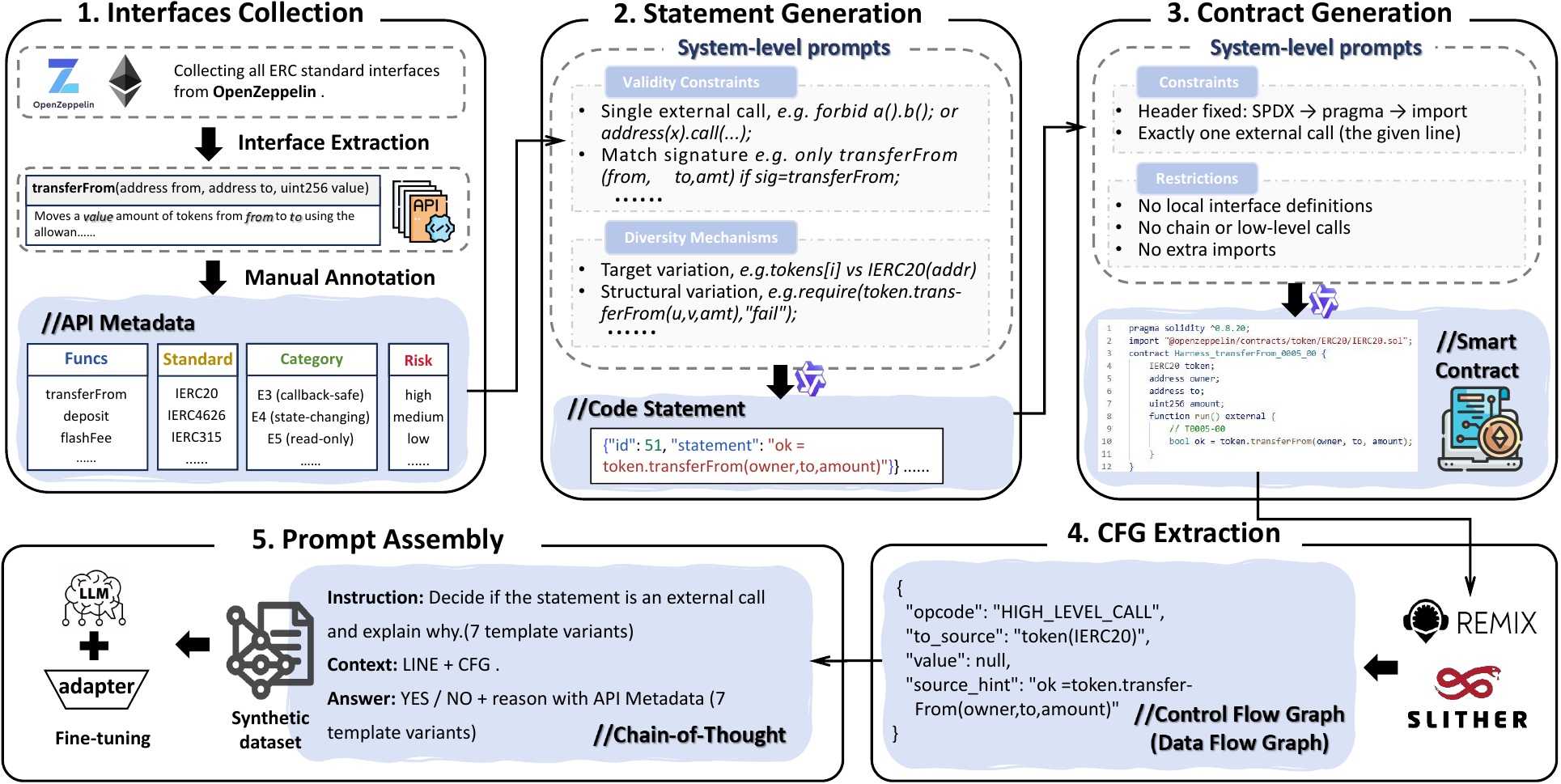}
  \caption{Pipeline of External-Call Synthetic Data Construction. (1) Interfaces are collected from OpenZeppelin and annotated with categories and risks. (2) Statements are generated under validity constraints to ensure correctness, with diversity enhancing structural and parametric coverage. (3) Statements are generated by rule into minimal compilable contracts. (4) Slither performs compilation checks and extracts CFG. (5) Data are assembled as instruction–answer templates for training.}
  \label{fig:external-call-pipeline}
\end{figure*}

\textit{Seed annotations.}
We begin with a small set of high-quality examples annotated by
domain experts. These examples are drawn from canonical ERC standards
(e.g., ERC20, ERC721, ERC777, ERC1155) and OpenZeppelin libraries,
where the semantics of external calls are unambiguous. They serve as
anchors for what constitutes a hijackable external call versus a
non-hijackable one.

\textit{Controlled synthesis.}
Building on these seeds, we systematically expand the dataset through
template-based generation and program mutation. The overall pipeline is illustrated in Figure~\ref{fig:external-call-pipeline}. Specifically:
\begin{itemize}[nosep]
  \item \emph{Interface coverage:} APIs from multiple ERC standards are
  enumerated to ensure diversity of call targets.
  \item \emph{Statement synthesis:} high-variance templates generate
  single-line call statements with varied invocation styles.
  \item \emph{Contract embedding:} each statement is placed into a
  minimal compilable contract skeleton to guarantee validity.
  \item \emph{CFG extraction:} static analysis tools (Slither, Remix)
  are used to extract control-flow context, which is included with
  probability 0.7 to augment the input.
  \item \emph{Instruction assembly:} each instance is packaged as an
  instruction–context–answer triple, aligned with instruction-tuning
  formats.
\end{itemize}

\textit{Preference refinement.}
The synthetic samples are then filtered and refined by consistency
checks and preference-based selection. Rule-based validators remove
contradictory or ill-formed cases, while preference optimization
prioritizes samples aligned with domain semantics, thereby enhancing
data reliability.

\subsection{Data-Dependency Dataset}
\label{sec:dataset-t3}
The data-dependency dataset determines whether an external call $e$ and a state update $s$ are semantically dependent. It integrates rule-driven synthesis with CFG-guided validation to balance positive and negative cases. Compared with external-call ’s interface-based generation, it requires path-sensitive reasoning and structured supervision, making the dependency factor $\phi_D$ both challenging and complementary to the external-call factor $\phi_E$. The construction of the sorting dataset is similar to this and see appendix C for details.

\textit{Challenges.}
Dependencies are not guaranteed by adjacency: $e$ and $s$ may be separated by arithmetic operations, conditionals, or temporary variables. Correctness depends on path feasibility, and dependency types are diverse, including direct parameter passing, return-value propagation, and slot-level addressing. To prevent shortcut learning, negative cases must closely resemble positives while breaking the actual flow.

\textit{Construction approach.}
We adopt a rule-driven synthesis strategy. 
A dependency rule table specifies direction (E$\to$S / S$\to$E), valid attachment points, and whether intermediate transformations are mandatory. Each contract contains exactly one \texttt{//e} and one \texttt{//s}, separated by at least one unrelated statement, ensuring both dependency and non-dependency variants. To verify semantic correctness, contracts are compiled into IR, a control-flow graph and def--use map are extracted, and we check whether variables at $s$ can be reached from the call frontier along some feasible path. Only validated pairs are retained.

\setlength{\tabcolsep}{3.5pt}      
\renewcommand{\arraystretch}{1.12} 

\begin{table*}[!t]
  \centering
  \Large
  \caption{Effectiveness of synthetic datasets under different factor settings.}
  \label{tab:synthetic-validity}
  \begin{adjustbox}{width=\textwidth}
  \begin{tabular}{@{} l l *{12}{c} @{}}
    \toprule
    \multirow{2}{*}{Model} & \multirow{2}{*}{Variant}
      & \multicolumn{4}{c}{E}
      & \multicolumn{4}{c}{D}
      & \multicolumn{4}{c}{O} \\
    \cmidrule(lr){3-6}\cmidrule(lr){7-10}\cmidrule(lr){11-14}
      & & P & R & F1 & ACC
        & P & R & F1 & ACC
        & P & R & F1 & ACC \\
    \midrule
    \multirow{3}{*}{%
      \begin{tabular}[t]{@{}l@{}}
        StableCode-3B
      \end{tabular}
    }
      & Base    & 37.44 & 61.72 & 46.61 & 44.65
               & 60.34 & 100.00 & 75.27 & 60.34
               & 6.63 & 25.00 & 10.49 & 26.54 \\
      & ZeroGen & 0.00 & 0.00 & 0.00 & 0.00
               & 100.00 & 0.71 & 1.42 & 40.09
               & 18.89 & 17.18 & 16.30 & 18.92 \\
      & LoRA    & 93.72 \sd{5.92} & 93.31 \sd{6.05} & 93.47 \sd{5.09} & 97.39 \sd{1.72}
               & 100.00 \sd{0.00} & 100.00 \sd{0.00} & 100.00 \sd{0.00} & 100.00 \sd{0.00}
               & 100.00 \sd{0.00} & 100.00 \sd{0.00} & 100.00 \sd{0.00} & 100.00 \sd{0.00} \\
    \midrule
    \multirow{3}{*}{%
      \begin{tabular}[t]{@{}l@{}}
        CodeLlama-7B
      \end{tabular}
    }
      & Base    & 26.92 & 5.47 & 9.09 & 57.19
               & 66.67 & 1.43 & 2.80 & 40.09
               & 26.92 & 5.47 & 9.09 & 57.19 \\
      & ZeroGen & 42.50 & 99.90 & 59.65 & 7.39
               & 67.44 & 20.71 & 31.69 & 46.12
               & 29.49 & 33.82 & 31.51 & 61.54 \\
      & LoRA    & 95.07 \sd{3.60} & 100.00 \sd{0.00} & 97.44 \sd{1.74} & 99.30 \sd{0.50}
               & 100.00 \sd{0.00} & 100.00 \sd{0.00} & 100.00 \sd{0.00} & 100.00 \sd{0.00}
               & 66.67 \sd{47.14} & 66.67 \sd{47.14} & 66.67 \sd{47.14} & 66.67 \sd{47.14} \\
    \midrule
    \multirow{3}{*}{%
      \begin{tabular}[t]{@{}l@{}}
        Qwen-14B
      \end{tabular}
    }
      & Base    & 42.86 & 86.72 & 57.36 & 49.54
               & 59.73 & 96.43 & 73.77 & 58.62
               & 25.68 & 97.06 & 40.62 & 25.77 \\
      & ZeroGen & 70.39 & 99.21 & 82.35 & 83.43
               & 74.81 & 72.14 & 73.45 & 68.53
               & 35.71 & 22.06 & 27.27 & 69.23 \\
      & LoRA    & 99.30 \sd{0.26} & 99.41 \sd{0.21} & 99.36 \sd{0.24} & 99.35 \sd{0.22}
               & 98.93 \sd{0.51} & 99.64 \sd{0.51} & 99.22 \sd{0.59} & 99.13 \sd{0.62}
               & 99.78 \sd{0.22} & 99.82 \sd{0.19} & 99.80 \sd{0.20} & 99.79 \sd{0.22} \\
    \bottomrule
  \end{tabular}
  \end{adjustbox}

\begin{tablenotes}
    \footnotesize
    \raggedright
    \item \textit{E, D, and O correspond to the three factor-specific settings evaluated in this table, representing External Call, Dependency, and Ordering, respectively.}
\end{tablenotes}

\end{table*}

\section{Experiments}

\subsection{Experimental Settings}
All experiments are built upon the \textit{Qwen2.5-Coder-14B-Instruct} model as the backbone.  All experiments are executed on two NVIDIA A100 (80GB) GPUs.  We compare CompFuse with both internal and external baselines.  Internal baselines include non-fusion, frozen-weight fusion, and trainable adaptive fusion variants built on the same Qwen2.5-Coder-14B backbone.  For external comparison, we benchmark against five representative reentrancy analyzers: Slither~\cite{feist2019slither}, Mythril~\cite{mythril2025}, Securify~\cite{tsankov2018securify}, Sailfish~\cite{rao2012sailfish}, and Smartian~\cite{choi2021smartian}, all executed with their default configurations.  

Our experiments involve both synthetic and real-world datasets.  
The synthetic datasets contain approximately 2,500 samples for each of three factor-level tasks—external call recognition (E), dependency identification (D), and ordering adjustment (O).

All samples are verified by Slither to ensure compilability and semantic soundness. The fusion experiments are trained on datasets summarized from four existing studies, containing 166 vulnerable and 800 non-vulnerable samples~\cite{yang2023definition,ferreira2020smartbugs,zheng2023turn,cai2025detecting}. We also curate a benchmark of 31 real-world Solidity contracts confirmed to contain reentrancy vulnerabilities through public incident reports and manual verification.  


\begin{figure}[!t]
  \centering
  \includegraphics[width=1.00\linewidth]{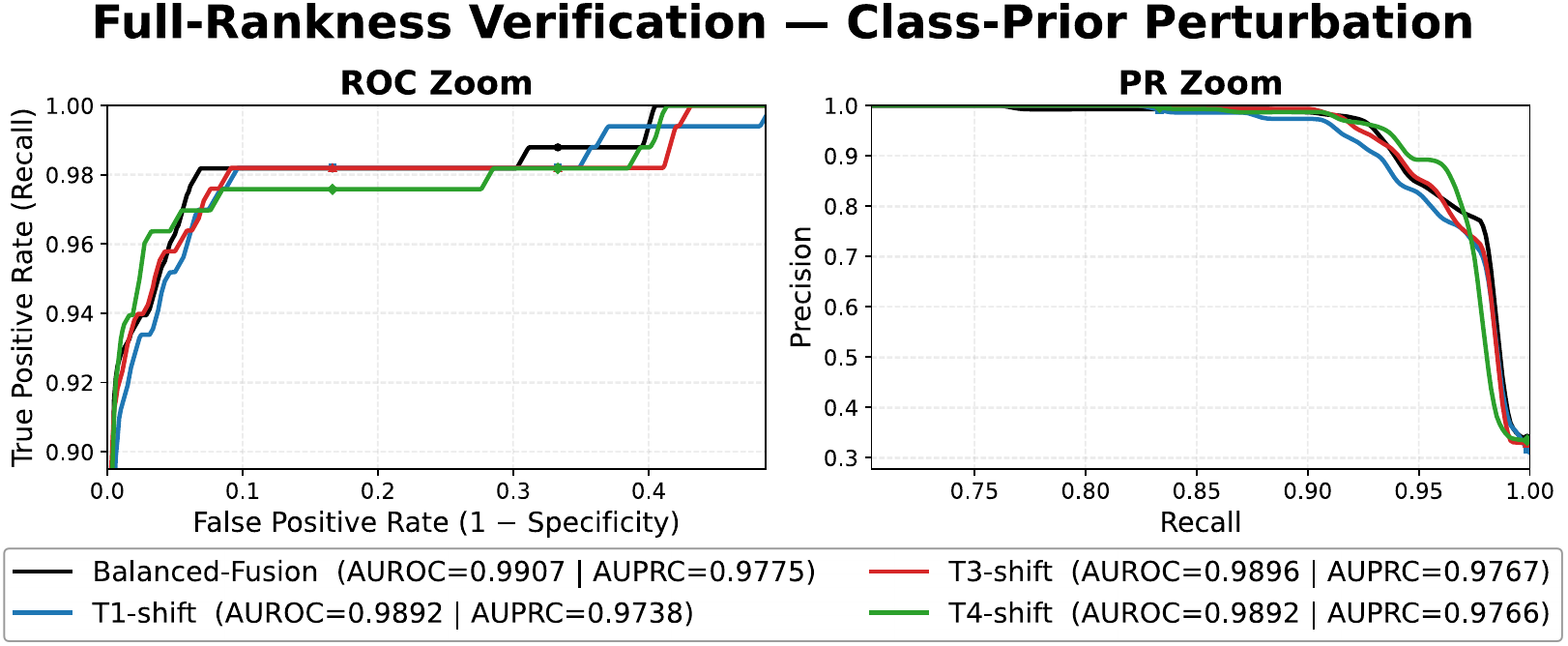}
  \caption{
    Full-Rankness Verification.}
  \label{fig:orthogonality}
\end{figure}

\subsection{Synthetic Dataset Validation}

This experiment evaluates whether compositional supervision can be induced through compiler-verified and graph-augmented synthetic datasets. 
Each dataset encodes factor-level supervision aligned with structural rules of control and data flow, allowing us to test whether LLMs can internalize dependencies among external calls, state updates, and execution order beyond surface memorization.  
We conduct experiments using three instruction-tuned code models: 
Stable-Code-Instruct-3B\footnote{\url{https://huggingface.co/stabilityai/stable-code-instruct-3b}, an open 3B model optimized for multilingual code synthesis.}, 
CodeLlama-7B-Instruct-Solidity\footnote{\url{https://huggingface.co/AlfredPros/CodeLlama-7b-Instruct-Solidity}, a Solidity-specialized variant of CodeLlama-7B.}, 
and Qwen2.5-Coder-14B-Instruct\footnote{\url{https://huggingface.co/Qwen/Qwen2.5-Coder-14B-Instruct}, a 14B model trained for advanced code reasoning and multi-language support.}, 
ensuring coverage across different parameter scales.

As shown in Table~\ref{tab:synthetic-validity}, LoRA-tuned models achieve consistent and near-saturated results across all tasks, with Qwen-14B exceeding 99\% F1 and accuracy.Given that the base model already achieves strong performance on the state-update factor (S) (F1 > 90\%), we omit additional factor-specific fine-tuning.

These improvements over Base and ZeroGen variants confirm that compiler verification and CFG/DFG augmentation inject explicit structural knowledge rather than lexical patterns. 
Performance gains are strongest in the dependency factor (D), while mild fluctuations in the ordering factor (O) suggest that ordering reasoning remains data-sensitive.

Larger models integrate multi-granular cues such as API semantics and control-flow dependencies more effectively, whereas smaller models show higher variance but still benefit from the structured prompts that enforce clear factor boundaries.  
\textbf{In summary, compiler-verified and graph-augmented datasets provide reliable compositional supervision that enables LLMs to internalize reentrancy-relevant reasoning patterns beyond handcrafted heuristics.}

\subsection{Full-Rankness Verification}
We only verify Definition~\ref{def:rankness} (Full-rankness) because the other three conditions are theoretically ensured by model design and data construction.
To examine whether the proposed compositional framework preserves balanced factor contributions when the supervision of a single factor is perturbed, we perform an auxiliary experiment aligned with the full-rankness assumption defined in Section~\ref{analysis} and formalized in Definition~\ref{def:rankness}.
Specifically, we modify the positive–negative ratios of tasks T1, T3, and T4 from 1:2 to 0.5:0.95, while keeping all other configurations identical to the main setup.
This adjustment introduces a controlled class-prior shift to test whether the adaptive fusion can maintain equilibrium among the four factors $(E,S,D,O)$ during training.
We record the AUROC trajectories of three independent fusion runs, each reflecting how the model integrates these factors over optimization steps.

As shown in Figure~\ref{fig:orthogonality}, the AUROC curves from the three runs almost perfectly overlap throughout training, with average deviations below 0.02 across both class-prior settings.
The model converges stably under the altered supervision, showing no sign of directional collapse or overfitting to any modified task distribution.
This consistent behavior indicates that the adaptive fusion continues to leverage all factor branches in a balanced manner, thereby maintaining full-rank compositional representations.
\textbf{In summary, the near-identical AUROC trajectories across different class-prior ratios empirically confirm the balanced contribution of all factors and validate the correctness of Definition~\ref{def:rankness} in Section~\ref{analysis}.}

\subsection{Fusion Experiments}

We evaluate five configurations on the same frozen backbone: a prompt-free baseline (Base-model), a zero-shot variant (Zero-Shot), a single-task model fine-tuned only on reentrancy data (Only-LoRA), a uniform fusion model that averages four factor adapters with fixed weights (Fusion-frozen, $\alpha_k{=}0.25$), and an adaptive fusion model with trainable gating weights (Fusion-train). 
The first two serve as conventional baselines, while the latter three progressively incorporate task-specific supervision and compositional fusion to assess how structured integration improves vulnerability detection.

As shown in Table~\ref{tab:fusion-variants}, fusion-based variants substantially outperform the non-fusion baselines, confirming that aggregating factor representations is crucial for reliable detection. 
Compared with Only-LoRA, the Fusion-frozen achieves a clear improvement in F1, increasing from 81.08 to 89.04, which demonstrates that orthogonal factor branches contribute complementary information. 
Further introducing adaptive gating in Fusion-train enhances both recall and overall accuracy, indicating that the model dynamically reweights factor importance according to contextual relevance. 
This adaptive mechanism allows the fused representation to emphasize decisive cues—such as execution order and dependency consistency—while suppressing redundant signals. \textbf{In summary, the adaptive fusion variant achieves the most balanced and interpretable composition, yielding more stable detection and stronger generalization to unseen contract structures.}

\begin{table}[t]
  \centering
  \normalsize   
  \setlength{\tabcolsep}{8pt}  
  \renewcommand{\arraystretch}{1.2}  
  \caption{Comparison of Fusion Variants for Reentrancy Detection.}
  \label{tab:fusion-variants}
  \resizebox{0.97\columnwidth}{!}{  
  \begin{tabular}{lcccc}
    \toprule
    \textbf{Variant} & {P} & {R} & {F1} & {ACC} \\
    \midrule
    Base-model     & 17.01 & 20.00 & 18.38 & 69.64 \\
    Zero-Gen      & 54.90 & 33.94 & 41.95 & 83.94 \\
    Only-Lora      & 78.95 & 83.33 & 81.08 & 92.86 \\
    Fusion-Frozen (Ours)  & \underline{\textbf{99.26}} & 80.72 & 89.04 & 96.58 \\
    \textbf{CompFuse (Ours)} & 98.06 & \underline{\textbf{91.57}} & \underline{\textbf{94.70}} & \underline{\textbf{98.24}} \\
    \bottomrule
  \end{tabular}}
\end{table}

\subsection{Ablation Study}

To evaluate the contribution of each factor branch, we conduct a controlled ablation by enabling different combinations of factor-specific adapters corresponding to the external-call (\textbf{E}), state-update (\textbf{S}), dependency (\textbf{D}), and ordering (\textbf{O}) factors, while keeping the fusion and gating modules intact.
, while keeping the fusion and gating modules intact. 
This design isolates the functional importance of each component within the trained fusion model and examines how individual factors influence compositional reasoning.

As shown in Table~\ref{tab:ablation}, removing any single branch results in a clear performance degradation, confirming that all factors are indispensable for reconstructing the full semantics of reentrancy. 
The largest drop occurs when the external-call adapter is removed, demonstrating that call-site identification serves as the primary anchor for reasoning about subsequent state updates and control flow. 
Disabling either the state-update or ordering adapters also weakens detection, indicating that both contribute complementary cues—one capturing mutation semantics and the other maintaining causal consistency across execution paths. 
Together, these patterns reveal that robust detection relies on the interplay rather than the dominance of any single factor. \textbf{In summary, the ablation confirms that the fusion model’s effectiveness stems from the synergistic interaction among compositional factors rather than isolated component performance.}

\begin{table}[t]
  \centering
  \normalsize
  \setlength{\tabcolsep}{8pt}
  \renewcommand{\arraystretch}{1.2}
  \caption{Ablation Study on Factor-specific Adapters.}
  \label{tab:ablation}
  \resizebox{0.97\columnwidth}{!}{
  \begin{tabular}{lcccc}
    \toprule
    \textbf{Enabled Factors} & \textbf{P} & \textbf{R} & \textbf{F1} & \textbf{ACC} \\
    \midrule
    D+O                  & 92.86 & 83.87 & 88.14 & 96.37 \\
    E+O                  & 87.10 & 93.10 & 90.00 & 96.89 \\
    E+D                  & \textbf{100.00} & 82.35 & 90.32 & 96.89 \\
    ComFuse (Full)       & 98.06 & \textbf{91.57} & \textbf{94.70} & \textbf{98.24} \\
    \bottomrule
  \end{tabular}}
\begin{tablenotes}
    \footnotesize
    \raggedright
    \item \textit{E, S, D, and O denote the External Call, State Update, Dependency, and Ordering factors, respectively. Combinations such as ``E+D'' indicate that only the specified factor-specific adapters are enabled. ``Full'' denotes using all four factors.}
\end{tablenotes}

\end{table}

\subsection{Comparison with Existing Tools}

We evaluate the proposed fusion model against several mainstream reentrancy analyzers on verified real-world contracts containing confirmed vulnerabilities. 
This comparison aims to examine whether compositional fusion provides complementary advantages to traditional symbolic and static analyses under realistic auditing conditions. 
All models are tested on identical contracts to ensure a fair and interpretable evaluation of practical detection capability.

As summarized in Table~\ref{tab:tool-compare-pos}, CompFuse achieves a notably higher recall (87.1\%) and detects more true vulnerabilities with fewer omissions. 
Conventional analyzers remain effective for canonical patterns but may be limited when contract structures involve complex control dependencies, ERC-standard interfaces, or nested invocation flows that diverge from predefined heuristics. 
The proposed model complements these methods by leveraging learned compositional representations that capture semantic relationships among external calls, state updates, and execution order. 
Such representations allow flexible generalization across unseen structural variants, leading to more stable and consistent detection in heterogeneous contract environments. \textbf{In summary, compositional fusion provides a complementary and generalizable approach to smart contract analysis, achieving robust reentrancy detection beyond the coverage of conventional rule-based tools.}

\begin{table}[t]
  \centering
  \caption{Comparison with Existing Reentrancy Analyzers.}
  \label{tab:tool-compare-pos}
  \resizebox{\linewidth}{!}{
  \begin{tabular}{l c c c c c c}
    \toprule
    Metric   & Securify & Sailfish & Smartian & Mythril & Slither & \makecell{\textbf{CompFuse}\\(Ours)} \\
    \midrule
    TP     & 4 & 4 & 5 & 6 & 19 & \underline{\textbf{27}} \\
    FN     & 27 & 27 & 26 & 25 & 11 & \underline{\textbf{4}} \\
    Recall (\%)   & 12.90 & 12.90 & 16.13 & 19.35 & 63.33 & \underline{\textbf{87.10}} \\
    \bottomrule
  \end{tabular}}
\begin{tablenotes}
    \footnotesize
    \raggedright
    \item \textit{This table reports recall only because the evaluation is conducted on a real-world benchmark consisting exclusively of contracts confirmed to contain reentrancy vulnerabilities, for which source code has been publicly disclosed in recent years. Under this setting, precision-based metrics are not applicable. Recall directly reflects a tool’s ability to identify known vulnerabilities without omission under realistic auditing conditions.}
\end{tablenotes}

\end{table}

\section{Discussion}
\label{sec:discussion}
This research shows that decomposing complex reasoning tasks into atomic subtasks and fusing their weights can improve accuracy in data-sparse vertical domains. By separately supervising factors such as external calls, state updates, dependencies and execution order, this approach enhances the model's understanding of reasoning rules, thereby improving robustness to unknown patterns. Compared to traditional vulnerability detection methods that rely on a single semantic feature or end-to-end training, our strategy offers new insights into task decomposition and rule understanding. This suggests that combinatorial generalization not only improves performance on specific tasks but also serves as an effective approach to teaching general rules for complex reasoning tasks. 

However, this research is limited by the size of the synthetic dataset and the complexity of rule design. Future work should extend this approach to more vulnerability types and real-world scenarios to further validate its robustness and generalization capabilities. Specifically, integer overflow vulnerabilities arise primarily due to the lack of effective bounds or overflow checks on variables involving external inputs in arithmetic operations before state updates. Therefore, integer overflow vulnerabilities can be decomposed into the following four factors: identifying all external input variables in the contract, identifying state update statements related to the input variables (i.e., data dependencies), identifying bounds or overflow check statements, and determining the execution order of state update statements and check statements. In general, this research provides a new framework CompFuse for interpretable modeling of complex reasoning tasks and offers potential directions for automated analysis in high-risk areas such as smart-contract vulnerability detection.

\section{Related work}
\textit{Out-of-distribution generalization}
The concept of out-of-distribution (OOD) generalization was first introduced by Bastings et al.~\cite{bastings2018jump}, who proposed semantic generalization tasks on synthetic datasets (e.g., walk twice → jump twice). They found that contemporary deep learning models such as RNNs and LSTMs were often limited to memorizing training mappings rather than achieving true generalization. Subsequent work extended the SCAN dataset to Transformer architectures to evaluate their generalization ability~\cite{ontanon2022making, csordas2002somatic}, and suggested that generalization failures may stem from model bias, where models tend to memorize mappings instead of learning underlying rules. To further explain this phenomenon, theoretical studies~\cite{soudry2018implicit, arora2018stronger} demonstrated that smaller weight norms are typically associated with stronger generalization, while higher model complexity can exacerbate memorization tendencies. Building on this line of inquiry, later research identified the neural collapse phenomenon~\cite{sukenik2024neural, zangrando2024neural}, and hypothesized that low-rank solutions may be connected to rule learning. Meanwhile, in continual learning settings, studies have shown that organizing tasks by difficulty or dependency order can enhance generalization~\cite{fu2024exploring}. More recently, Zhang et al.~\cite{zhang2025complexity} systematically incorporated initialization scale, weight decay, and stable rank into a unified framework for compositional OOD generalization, further advancing the study of generalization beyond the training distribution.

\textit{Reentrancy vulnerabilities}
Reentrancy is one of the earliest discovered vulnerabilities in smart contracts. The well-known DAO incident~\cite{mehar2019understanding} exploited external calls to repeatedly enter the withdrawal logic before the contract updated its internal state, thereby draining funds. Early approaches primarily relied on symbolic execution and dependency-graph–based static analyses to formalize and detect reentrancy vulnerabilities~\cite{luu2016making, tsankov2018securify, feist2019slither, xue2020cross}. However, static tools are often limited in capturing vulnerabilities that arise from runtime transaction sequences, which motivated later work to explore dynamic detection and runtime protection mechanisms~\cite{rodler2018sereum, so2021smartest}. A common theme across these studies—especially in static analyses—is to treat the ordering of external calls and state updates as the core condition for identifying reentrancy~\cite{luu2016making, tsankov2018securify, rodler2018sereum, xue2020cross, so2021smartest}. Consequently, recombining these elements of external calls, state updates, and their ordering provides a natural foundation for compositional generalization.

\section{Conclusion}
This work addresses the limited compositional generalization of large language models on out-of-distribution reasoning tasks. We propose a hybrid approach that decomposes complex reasoning into atomic subtasks and recombines them to enhance the model’s understanding of complex logic. To validate the approach, we model reentrancy vulnerabilities in smart contracts and construct synthetic datasets for training with low-rank adaptation. Experimental results demonstrate that our method achieves higher accuracy and stronger generalization compared with baseline models.

\bibliographystyle{ACM-Reference-Format}
\bibliography{sample-base}

\appendix
\renewcommand{\theassumption}{\arabic{assumption}} 
\renewcommand{\thetheorem}{\arabic{theorem}}

\section*{A. Algorithm of fusion}

\begin{algorithm}[t]
\caption{Training with Gated Fusion}
\label{alg:cjmoa}
\begin{algorithmic}[1]
\Require Dataset $\mathcal{D}$; frozen backbone $f_{\theta_0}$; adapters $\{E,S,D,O\}$;
enable mask $m$; prior $\pi$; temperature $\tau$; Jacobian weight $\lambda_{\text{jaco}}$;
(optional) frozen weights $\bar{\alpha}$; threshold $\delta$; gate warm-up ratio $\rho$
\Ensure Trained gating parameters and classifier head
\For{each epoch}
  \For{each batch $(x,y)$ from $\mathcal{D}$}
    \State \textbf{Feature extraction:} obtain $\theta_E(x),\theta_S(x),\theta_D(x),\theta_O(x)$ 
    \If{$\bar{\alpha}$ exists}
        \State $\alpha \gets \bar{\alpha}$ 
    \Else
        \State Compute channel scores $s_k(x)$ for $k\in\{E,S,D,O\}$
        \State Apply mask: $s_k(x)\!\gets\!-\infty$ if $m_k=0$
        \State $\alpha^{\text{on}}_k(x)\!\gets\!\dfrac{\exp(s_k(x)/\tau)}{\sum_j \exp(s_j(x)/\tau)}$
        \State $\alpha_k(x)\!\gets\!\dfrac{\alpha^{\text{on}}_k(x)\,m_k}{\sum_j \alpha^{\text{on}}_j(x)\,m_j}$
    \EndIf
    \State \textbf{Fusion:} $Z(x)\!\gets\!\sum_{k\in\{E,S,D,O\}} \alpha_k(x)\,\theta_k(x)$
    \State \textbf{Prediction:} $z(x)\!\gets\!\langle w, Z(x)\rangle+b$,\quad $\hat{y}(x)\!=\!\sigma(z(x))$
    \State \textbf{Cross-entropy:} $L_{\mathrm{CE}}\!\gets\!\mathrm{CE}(y,\hat{y}(x))$
    \If{no fixed $\bar{\alpha}$ and $\lambda_{\text{jaco}}>0$}
        \State Let $z_{y^*}(x)$ be the logit of the ground-truth class
        \State Gradient norms $g_k\!=\!\big\|\partial z_{y^*}(x)/\partial \theta_k(x)\big\|_2$
        \State Normalize $q_k\!=\!g_k/\sum_j g_j$ 
        \State Active-set gate $\hat{\alpha}_k\!=\!\alpha_k/\sum_j \alpha_j m_j$
        \State $L_{\text{jaco}}\!\gets\!\mathrm{KL}\!\big(q\,\|\,\hat{\alpha}\big)$
    \Else
        \State $L_{\text{jaco}}\!\gets\!0$
    \EndIf
    \State \textbf{Total loss:} $L\!\gets\!L_{\mathrm{CE}}+\lambda_{\text{jaco}}\,L_{\text{jaco}}$
    \State \textbf{Optimization:}
      \If{current step $< \lceil \rho\cdot$ (total steps)$\rceil$}
        \State update only gating score parameters
      \Else
        \State update gating and classifier head parameters
      \EndIf
  \EndFor
\EndFor
\end{algorithmic}
\end{algorithm}

\subsection*{B.1 The proof of Theorem 3.1}

\begin{proof}
\textit{Gradient of $f$.} Let $P=P(x,\tilde z)$ be the program instance controlled by compositional factors $\tilde z\in[0,1]^4$ (corresponding to $E,S,D,O$), and define
\[
A_{ij}(P)\;=\;\phi_E(P)[i]\;\phi_S(P)[j]\;\phi_D(P)[i,j]\;\tilde{\phi}_O(P)[i,j].
\]
The scoring kernel in Eq.~\ref{soft} is
\[
f(P)\;=\;\log\!\sum_{i,j}\exp\!\big(A_{ij}(P)\big).
\]
Then,
\begin{equation}
\nabla_P f(P)
= \sum_{i,j}\mathrm{softmax}(A)_{ij}\,
  \nabla_P A_{ij}(P)
\end{equation}
where $\mathrm{softmax}(A)_{ij}
= \frac{e^{A_{ij}(P)}}{\sum_{a,b}e^{A_{ab}(P)}}$. Since each $\phi_k$ is Lipschitz, their Jacobians are bounded by $L_\phi$,
thus $\|\nabla_P A_{ij}(P)\|\le 4L_\phi$ and 
$\|\nabla_P f(P)\|\le L_f < \infty$ for some constant $L_f$.

\medskip
\textit{Chain rule to factor space.} Let $\tilde z=\sigma(\beta u) \subseteq (0,1)^4$ be the continuous relaxation 
of discrete factors with $\sigma(t)=1/(1+e^{-t})$ 
and temperature $\beta>0$.
Then
\[
J_{\tilde z}(u)
= \nabla_u \tilde z(u)
= \mathrm{diag}\!\big(\beta\,\sigma(\beta u_k)(1-\sigma(\beta u_k))\big),
\;
\|J_{\tilde z}(u)\|\le \frac{\beta}{4}.
\]
Applying the chain rule,
\begin{equation}
\nabla_{u} f(P(x,\tilde z(u)))
= \nabla_{P}f(P)\,
  \nabla_{\tilde z}P(x,\tilde z)\,
  J_{\tilde z}(u).
\end{equation}
Taking norms and using the Lipschitz bounds gives
\begin{equation}
\big\|\nabla_{u} f(P(x,\tilde z(u)))\big\|
\le L_f\,L_P\,\frac{\beta}{4}.
\end{equation}
Thus the function is differentiable in $u$ (and hence in each $\tilde z_k$) 
and has a finite Lipschitz constant $L_f L_P \beta/4$.

\medskip
Since $f$ and $\phi_{E,S,D,O}$ are Lipschitz and differentiable, 
and the composition with the smooth relaxation $\tilde z=\sigma(\beta u)$ preserves differentiability, the overall scoring function $\tilde z\mapsto f(P(x,\tilde z))$ is smooth with bounded gradient norm. Therefore, the smoothness assumption defined above is satisfied.
\end{proof}

\subsection*{B.2 The proof of Theorem 3.2}

\begin{assumption}
The synthetic data generator introduces no unseen primitives at test time; 
for every $x\in\mathrm{supp}(Q)$, there exist $f_E,f_S,f_D,f_O\in\mathrm{supp}(P)$ such that $x=g(f_E,f_S,f_D,f_O)$.
\label{asu1}
\end{assumption}

\begin{assumption}
The fusion model in Eq.~\ref{eq:h} combines fixed adapters $\theta_k$ without modifying their internal parameters.
\label{asu2}
\end{assumption}

\begin{theorem}[Compositionality by Design]
Under Assumptions~\ref{asu1} and~\ref{asu2}, the compositionality condition  is guaranteed.
\end{theorem}

\begin{proof}
The synthetic data generator in Assumptions~\ref{asu1} guarantees that no new primitives are introduced at test time, ensuring that every test composition can be expressed as $x=g(f_E,f_S,f_D,f_O)$ for some $f_E,f_S,f_D,f_O\in\mathrm{supp}(P)$. Therefore, $x\in f(P)$ by definition. 
Furthermore, the fusion model uses adapters $\theta_k$ trained solely on training primitives and keeps them frozen, so $\Theta$ preserves the same compositional closure. Together, these two mechanisms ensure that both the data generation process and the model fusion respect the compositionality assumption.
\end{proof}

\subsection*{B.3 The proof of Theorem 3.3}
\begin{proof}
\textit{Dataset-level rank.}
Let each sample have a binary factor vector $z=(E,S,D,O)\in\{0,1\}^4$, and form the design matrix $Z=[z_1^\top;\dots;z_N^\top]\in\mathbb{R}^{N\times4}$. By construction, the dataset contains an anchor $z^{(0)}$ and four single-bit perturbations $\{z^{(0)}\!\oplus\! e_k\}_{k\in\{E,S,D,O\}}$, yielding the submatrix
\[
\widetilde Z=\begin{bmatrix}
(z^{(0)})^\top,
\{z^{(0)}\oplus \phi_k\}^\top_{k\in\{E,S,D,O\}}
\end{bmatrix}^\top
\]
satisfies $\widetilde Z-\mathbf{1}(z^{(0)})^\top=\begin{bmatrix}
0,\phi_E^\top,\phi_S^\top,\phi_D^\top,\phi_O^\top
\end{bmatrix}^\top$,
whose column space spans $\mathbb{R}^4$.
Therefore, $\mathrm{rank}(Z)=4$ and the empirical second moment $\widehat M=\frac{1}{N}\sum_i z_i z_i^\top$ is positive definite, ensuring non-degenerate factor coverage.

\textit{Representation-level separation.}
Let the adapter outputs be concatenated as 
$\Theta(x)=[\theta_E(x)^\top,\theta_S(x)^\top,\theta_D(x)^\top,\theta_O(x)^\top]^\top\in\mathbb{R}^{4d}$,
and define the fusion score as
\begin{equation}\label{eq:fusion-score}
s(x)=\langle w(x),\theta(x)\rangle,\; 
w(x)=G(\alpha(x)).
\end{equation}
Each $\theta_k$ is trained on a distinct atomic task and remains frozen during fusion, while $G$ maps the gating vector $\alpha$ to fusion weights.
Given the program construction $P(x,z)$ that toggles atomic factors, the overall mapping is $x\mapsto r(P(x,z))\mapsto s(x)$.

\textit{Jacobian with respect to factors.}
Relax $z$ to a continuous variable $\tilde z\in[0,1]^4$ for differentiation.
By the chain rule,
\begin{equation}\label{eq:jac}
J(x)=\nabla_{\tilde z}s(x,\tilde z)
=\nabla_{r}s(x)\,\nabla_{P}\Theta(P)\,\nabla_{\tilde z}P(x,\tilde z)
=\nabla_{r}s(x)\,\nabla_{P}\Theta(P)\,D(\tilde z)
\end{equation}
Since each atomic factor modifies a disjoint component of the program, $\nabla_{\tilde z}P$ is diagonal almost everywhere:
\[
D(\tilde z)=\mathrm{diag}(\delta_E,\delta_S,\delta_D,\delta_O),\quad \delta_k\neq0.
\]
The frozen adapters produce block-separable responses such that the four columns 
$v_k=\nabla_{r}s\,\nabla_{P}\Theta\,\phi_k$ are linearly independent.
Hence
\[
J(x)=\big[\delta_E v_E\;\;\delta_S v_S\;\;\delta_D v_D\;\;\delta_O v_O\big],\quad
\mathrm{rank}\,J(x)=4.
\]

\textit{Label balance as variance lower bound.}
Let $y\in\{0,1\}$ and assume the dataset matches realistic factor frequencies with $\Pr(E{=}1)$, $\Pr(S{=}1)$, $\Pr(D{=}1)$, $\Pr(O{=}1)\in[\varepsilon,1-\varepsilon]$ for some $\varepsilon>0$.
Then the empirical covariance
\[
\widehat\Sigma_{zz}=\frac{1}{N}\sum_i (z_i-\bar z)(z_i-\bar z)^\top
\succeq \varepsilon(1-\varepsilon) I_4,
\]
which ensures a positive variance lower bound and prevents degeneracy in factor space.

In summary, the dataset is non-degenerate in the latent factor space, the model representations are factor-separable, and the Jacobian with respect to $z$ is full rank almost everywhere. Therefore, the full-rankness condition holds.
\end{proof}

\subsection*{B.4 The proof of Theorem 3.4}

\begin{proof}
\textit{Same-domain requirement.}
Assumption requires training and testing to lie in the same semantic domain, i.e., $\mathrm{supp}(Q)\subseteq f(P)\subseteq\mathcal{S}$, so no new latent dimensions emerge at test time.
Equivalently, for any test composition $z^\star$, there exists a training anchor $z^{(0)}$ and a continuous path
\[
\gamma:[0,1]\to[0,1]^4,\; 
\gamma(0)=z^{(0)},\; \gamma(1)=z^\star,
\]
such that $P(x,\gamma(t))\in\mathcal{S}$ for all $t\in[0,1]$.

\textit{Continuity of the fused predictor.}
Since $G$ and each frozen adapter $\theta_k$ are continuous (Lipschitz) in their inputs, both $\Theta(\cdot)$ and $w(\cdot)$ are continuous.
Consequently, $s(x,z)$ is continuous in $z$.
Along the path $\gamma$, by the fundamental theorem of calculus:
\begin{equation}\label{eq:path-int}
s\big(x,z^\star\big)-s\big(x,z^{(0)}\big)
=\int_{0}^{1}\!\nabla_{z}s\big(x,\gamma(t)\big)\cdot \gamma'(t)\,dt.
\end{equation}
If $s$ is $L_s$-Lipschitz in $z$, then
\begin{equation}\label{eq:lipschitz-bound}
\big|\,s(x,z^\star)-s(x,z^{(0)})\,\big|
\le L_s\!\int_{0}^{1}\!\|\gamma'(t)\|\,dt
= L_s\,\mathrm{Len}(\gamma),
\end{equation}
which shows that $s$ admits a smooth extension from the training anchor $z^{(0)}$ to the test composition $z^\star$ within the same domain.

\textit{Task boundaries and semantic continuity.}
The modeling process enforces atomic boundaries via four projections:
\[
\Pi_E,\Pi_S,\Pi_D,\Pi_O:\ \mathcal{S}\to\mathcal{H},\;
\theta_k\circ P=\Pi_k\circ P,
\]
so that each sub-task controls a distinct semantic direction in $z$.
Because the fusion model combines only these task-specific frozen representations,
\[
\Theta(x)=F\!\big(\Pi_E(P),\Pi_S(P),\Pi_D(P),\Pi_O(P);\ \alpha(x)\big),
\]
the predictor respects identical semantic partitions in both training and testing.

In summary, this guarantees continuity of the learned mapping over the support domain and ensures a smooth functional extension from training to test regions. Hence, the consistency assumption is satisfied.
\end{proof}

\subsection*{C. Prompt Design of Datasets}

\begin{table*}[t]
\centering
\caption{Data Dependency Types and Constraints}
\label{tab:t3-depen}
\setlength{\tabcolsep}{4pt}
\renewcommand{\arraystretch}{1.10}
\footnotesize   
\begin{tabularx}{\textwidth}{l l Y p{7cm}}
\toprule
\textbf{Id} & \textbf{Name} & \textbf{Definition} & \textbf{Border} \\
\midrule
A\_DIRECT & Direct Dependency & s directly uses e’s input or return value without any non-identity transformation. &
- s must include the input or return value of e with identical variable names; no arithmetic / logical / hash transformation is allowed; if \texttt{returns\_type=void}, usage of return value is forbidden. \\
\addlinespace[2pt]
B\_INDIRECT & Indirect Dependency & s uses e’s input or return value, but only after computation, intermediate variable passing, or control flow propagation. &
- must involve non-identity transformation or intermediate step; if control-flow dependency: return value only appears in \texttt{if}/\texttt{?:} condition, RHS independent from e. \\
\addlinespace[2pt]
C\_CTRL & Control Dependency & execution of s fully depends on e’s return value, while the written RHS value itself is unrelated to e. &
- e’s return value only appears in the condition expression; the RHS and index of s are independent from e. \\
\addlinespace[2pt]
Z\_NONE & No Dependency & e and s are located in the same function but independent of each other. &
- s must not use e’s input or return value; execution of s must not be controlled by e. \\
\bottomrule
\end{tabularx}
\vspace{2pt}
\begin{flushleft}
\small
This table defines four \textbf{data dependency categories} for the \textit{dependency recognition} task (T3).
“Definition” specifies the semantic relationship between an external call ($e$) and a state update ($s$),
while “Border” provides formal constraints distinguishing valid and invalid cases.
The four dependency types—Direct, Indirect, Control, and None—capture how data flows connect external interactions to state mutations.
\end{flushleft}

\end{table*}

\begin{table*}[t]
\centering
\caption{CEI Behavior Taxonomy}
\label{tab:t4-cei-trimmed}
\setlength{\tabcolsep}{6pt}
\renewcommand{\arraystretch}{1.12}
\resizebox{\textwidth}{!}{%
\begin{tabular}{l l l l}
\toprule
\textbf{Label} & \textbf{Type} & \textbf{Ordering Requirement} & \textbf{Corresponding CEI Pattern} \\
\midrule
GOOD & CEI\_OK & Check $\rightarrow$ State Write $\rightarrow$ External Interaction. 
\texttt{require(...)} $\rightarrow$ \texttt{//s} $\rightarrow$ \texttt{//e} &
CEI \\
\midrule
RISK & SIMPLE\_INT\_BEFORE\_EFFECT & Interaction occurs before first write (unprotected). 
\texttt{//e → //s} &
(C)IE \\
\midrule
RISK & POST\_INTERACTION\_EFFECTS & A write still occurs after interaction. 
\texttt{//s → //e → //s} &
CEIE \\
\midrule
RISK & PATH\_SENSITIVE\_I\_BEFORE\_E &
\makecell[l]{On some paths, the interaction precedes the write (if / try / catch / early return).\\
A certain branch executes \texttt{//e → //s}} &
Path-Sensitive CEI Variant \\
\bottomrule

\end{tabular}
}
\vspace{2pt}
\begin{flushleft}
\small
This table categorizes typical \textbf{Checks–Effects–Interactions (CEI) orderings} for the \textit{CEI-principle detection} task (T4).
Each label describes a structural execution pattern between state writes (“s”) and external interactions (“e”),
indicating whether the contract complies with secure CEI ordering.
“GOOD” entries correspond to correct CEI adherence, whereas “RISK” types denote common reentrancy-prone sequences.
\end{flushleft}
\end{table*}

Stage I: External Call Generation
\paragraph{}{Step 1: Line-level External Call Generation}


\begin{tcolorbox}[enhanced, breakable, floatplacement=tbp,
  before skip=6pt, after skip=6pt, width=\columnwidth,
  colback=gray!5!white, colframe=gray!40!black, boxrule=0.3pt,
  arc=0.8mm, left=1mm,right=1mm,top=1mm,bottom=1mm,
  title=System Prompt,
  coltitle=white]
  
You are a senior Solidity assistant. Given a set of existing single-line external call statements,
continue generating \textbf{new, deduplicated, and structurally diverse} statements.
The output is used for external call recognition and CFG extraction.

\textbf{Hard Constraints:}
\begin{itemize}
  \item Output \textbf{JSON only}: \{\texttt{"lines": ["...;", "...;"]}\}, with length exactly $N$.
  \item Each element must be a single-line Solidity statement ending with a semicolon.
  \item Each line contains \textbf{exactly one external call}, whose function name and parameters match the given \texttt{sig}.
  \item Disallow: chained calls, secondary external calls, \texttt{this.x()}, \texttt{address(...).call/delegatecall/staticcall}, \texttt{try/catch}, or \texttt{send/transfer}.
  \item If the function is read-only (e.g., \texttt{balanceOf}, \texttt{name}), it must not modify state, emit events, or increment counters.
\end{itemize}

\textbf{Diversity Criteria:}
\begin{itemize}
  \item New outputs must be distinct from all existing lines both in string and structure.
  \item Structural diversity should differ in at least:
  \begin{enumerate}
      \item Target instance form (interface variable, mapping, struct field, casted address)
      \item Contextual structure (assignment, require, if-block, inline block, event arg, etc.)
      \item Parameter source or mathematical form (scaling, modulo, type-cast, hash mix, etc.)
  \end{enumerate}
  \item Variable name change alone is insufficient; diversity must reflect distinct structure or semantics.
\end{itemize}

\textbf{Randomization Strategy:}
Sampling should vary across combinations of structure, instance, and parameter style, guided by a variant index seed.

\textbf{Variable Pools:}
Token-like: \texttt{token, asset, base, lpToken, collateral, debt}\\
Receiver-like: \texttt{to, recipient, vaultAddr, sink}\\
Operator-like: \texttt{owner, spender, operator, user}\\
Amount-like: \texttt{amount, value, shares, liquidity}\\
Misc.: \texttt{idx, key, tag, flag, nonce, salt}

\textbf{Target Instance Library:}
Interface variable call (\texttt{token.f(...)}), mapping or array element (\texttt{assets[user].f(...)}),
nested struct access (\texttt{vaults[id].info.asset.f(...)}), interface casting (\texttt{Standard(addr).f(...)}),
or constant instance (\texttt{PRIMARY.f(...)}).

\textbf{Parameter Source Library:}
Direct variables, arithmetic expressions, scaling, modulo, bitwise ops, min/max, type casts, hash encodings, or indexed values.

\textbf{Context Structure Library:}
Assignment, \texttt{require}, single-line \texttt{if}, ternary form, inline block, \texttt{unchecked\{\}},
event argument, inner function argument, boolean test, or index write.

\textbf{Output Validation:}
Ensure exactly one external call per line, correct function signature, and JSON output format.
\end{tcolorbox}

\paragraph{}{Step 2: Contract-level Minimal Compilation Generation}
\begin{tcolorbox}[title=System Prompt,
  coltitle=white]
You are a senior Solidity engineer. Given a single external call line and its interface metadata,
generate a \textbf{minimal compilable contract} suitable for CFG extraction.

\textbf{Hard Rules:}
\begin{itemize}
  \item Output only complete Solidity source code, without explanations or Markdown.
  \item The first three lines must be:
  \begin{enumerate}
    \item \texttt{// SPDX-License-Identifier: MIT}
    \item \verb|pragma solidity ^0.8.20;|
    \item The exact import statement (e.g., \texttt{import "@openzeppelin/contracts/token/ERC721/\\ERC721.sol";})
  \end{enumerate}
  \item Do not declare any interface locally; all types must come from the imported file.
  \item Exactly one external call: the provided line, inserted immediately below the comment \texttt{// <label\_id>}.
  \item Disallow any additional external calls, low-level calls, or ETH transfer syntax.
\end{itemize}

\textbf{Type and Variable Preparation:}
Declare variables and data structures so that the given line compiles (e.g., declare \texttt{IERC20 token;} or \texttt{address addr;} depending on syntax).

\textbf{Wrapper Function (default: \texttt{run()})}
Mark as \texttt{nonpayable} by default; use \texttt{view} only for read-only calls;
use \texttt{payable} if \texttt{\{value: v\}} exists and target is payable.

\textbf{Output Requirements:}
Ensure the contract starts with the 3-line header, contains one public function with the labeled call,
compiles successfully, and has no redundant external calls.
\end{tcolorbox}

Stage II: Dependency-Constrained Contract Synthesis
\paragraph{}{Step 1: Base Contract Skeleton Generation}

\begin{tcolorbox}[enhanced, breakable, floatplacement=tbp,
  before skip=6pt, after skip=6pt, width=\columnwidth,
  colback=gray!5!white, colframe=gray!40!black, boxrule=0.3pt,
  arc=0.8mm, left=1mm,right=1mm,top=1mm,bottom=1mm,
  title=System Prompt,
  coltitle=white]

You are a Solidity contract generator. Given a sampled external call specification (E\_CALL), generate a \textbf{standalone, minimal, and directly compilable} single-file contract. The following hard requirements must be strictly satisfied:

\textbf{Header Constraints}
\begin{itemize}
  \item The first two lines must be fixed and appear in this exact order:
  \begin{verbatim}
// SPDX-License-Identifier: MIT
pragma solidity ^0.8.20;
  \end{verbatim}
\end{itemize}

\textbf{Import Rules}
\begin{itemize}
  \item Import is allowed \textbf{only if} \texttt{E\_CALL.import\_path} is non-empty.
  \item The import path must exactly match \texttt{E\_CALL.import\_path}.
  \item No additional imports, libraries, or dependencies are allowed.
  \item Contract inheritance is forbidden.
\end{itemize}

\textbf{Structural Constraints}
\begin{itemize}
  \item Generate exactly one contract.
  \item Generate exactly one externally visible function (\texttt{external} or \texttt{public}).
  \item The following constructs are forbidden:
  \begin{itemize}
    \item \texttt{assembly}
    \item \texttt{unchecked}
    \item \texttt{try/catch}
    \item \texttt{receive}
    \item \texttt{fallback}
    \item \texttt{modifier}
    \item \texttt{struct}
  \end{itemize}
\end{itemize}

\textbf{Unique External Call (e)}
\begin{itemize}
  \item There must be exactly one external call statement.
  \item The call must use the interface function specified by E\_CALL.
  \item The line immediately preceding the call must be the comment: \texttt{//e}
  \item If the return type is not \texttt{void}, the return value must be captured.
  \item No second external call is allowed.
\end{itemize}

\textbf{Unique State Update (s)}
\begin{itemize}
  \item There must be exactly one local state update statement.
  \item The state update must have a data dependency with the external call.
  \item The line immediately preceding the state update must be the comment: \texttt{//s}
  \item Only local state updates are allowed (no external calls or internal-call-based writes).
  \item No second state update is allowed.
\end{itemize}

\textbf{Dependency Enforcement}
\begin{itemize}
  \item The dependency between \texttt{//e} and \texttt{//s} must strictly follow the template specified by \texttt{DEP\_RULE.dep\_id}.
  \item Mixing dependency templates is forbidden.
  \item Direction and ordering are provided by the user instruction and must be strictly followed.
\end{itemize}

\textbf{Adjacency and Length Constraints}
\begin{itemize}
  \item The external call (e) and the state update (s) must not be adjacent.
  \item At least one unrelated statement must be placed between them.
  \item The total file length should be between 8 and 20 lines.
\end{itemize}

\textbf{Compilation Safety}
\begin{itemize}
  \item The contract must be directly compilable.
  \item Isolated type declarations (e.g., \texttt{uint256;}) are forbidden.
  \item Dynamic arrays must be allocated with \texttt{new} and filled element-wise.
  \item Assigning literal arrays to \texttt{uint256[]} is forbidden.
  \item Parameter and assignment types must match exactly.
  \item Implicit casts and fixed-to-dynamic array conversions are forbidden.
\end{itemize}

\textbf{Output Rules}
\begin{itemize}
  \item Output Solidity code only.
  \item No comments other than \texttt{//e} and \texttt{//s} are allowed.
  \item No empty lines.
  \item No explanatory text.
\end{itemize}

\textbf{External Instantiation Safety Guards}

\begin{itemize}
  \item Any external contract instantiation (e.g., \texttt{IERC20(...)}, \texttt{IERC721(...)}) must use a function parameter of type \texttt{address}.
  \item Hard-coded literals, constants, \texttt{msg.sender}, \texttt{tx.origin}, \texttt{this}, or expression chains are forbidden.
  \item Instantiating implementation contracts (e.g., \texttt{ERC20(...)}) is forbidden.
  \item If the interface requires an external address but no such parameter exists, a new address parameter must be added as the first function argument with semantic naming:
  \begin{itemize}
    \item IERC20 $\rightarrow$ \texttt{token}
    \item IERC721 $\rightarrow$ \texttt{nft}
    \item IERC1155 $\rightarrow$ \texttt{token1155}
    \item IERC1363 $\rightarrow$ \texttt{token1363}
    \item IERC3156FlashLender $\rightarrow$ \texttt{lender}
    \item Others $\rightarrow$ \texttt{target}
  \end{itemize}
\end{itemize}

\textbf{Allowed Instantiation Forms}
\begin{itemize}
  \item Inline call: \texttt{IERC20(token).transfer(to, amount);}
  \item Local binding: \texttt{IERC721 c = IERC721(nft); c.safeTransferFrom(...);}
\end{itemize}

\textbf{Interface and Import Binding}
\begin{itemize}
  \item Import is allowed only if \texttt{E\_CALL.import\_path} is non-empty.
  \item The imported type name must exactly match the file base name.
  \item No additional imports or guessed dependencies are allowed.
\end{itemize}

\textbf{Type Safety}
\begin{itemize}
  \item Casting integers to address or contract types is forbidden.
  \item Contract/interface $\rightarrow$ address requires explicit cast: \texttt{address(x)}.
  \item Implicit conversions are forbidden.
\end{itemize}

\textbf{Comment and Structure Rules}
\begin{itemize}
  \item Only two comments are allowed: \texttt{//e} and \texttt{//s}.
  \item If the state update is a subtraction and the rule requires protection, a \texttt{require(...)} must be placed before \texttt{//s}.
\end{itemize}

\end{tcolorbox}

\paragraph{}{Step 2: Per-Sample Dependency Constraint Injection}

\begin{tcolorbox}[enhanced, breakable, floatplacement=tbp,
  before skip=6pt, after skip=6pt, width=\columnwidth,
  colback=gray!5!white, colframe=gray!40!black, boxrule=0.3pt,
  arc=0.8mm, left=1mm,right=1mm,top=1mm,bottom=1mm,
  title=User Prompt,
  coltitle=white]

\textbf{E\_CALL (External Call Specification)}
\begin{itemize}
  \item Function signature: \texttt{\$\{e\_sig\}}
  \item Return type: \texttt{\$\{e\_ret\}}
  \item Interface name: \texttt{\$\{e\_ifn\}}
  \item Import path: \texttt{\$\{e\_imp\}}
\end{itemize}

If the return type is \texttt{void}, the return value must not be used as a dependency source. In this case, only the input parameters of the external call may be used to establish dependencies.

\textbf{S\_RULE (Local State Update Template)}
\begin{itemize}
  \item Rule ID: \texttt{\$\{s\_id\}}
  \item Name: \texttt{\$\{s\_name\}}
  \item Storage declaration: \texttt{\$\{s\_decl\}}
  \item Parameter specification: \texttt{\$\{s\_param\}}
  \item Write example: \texttt{\$\{s\_stmt\}}
  \item RHS type: \texttt{\$\{s\_rhs\}}
  \item Index arity: \texttt{\$\{s\_idx\}}
  \item Control flow: \texttt{\$\{s\_ctrl\}}
  \item Condition expression: \texttt{\$\{s\_cond\}}
  \item Dependency affordance: \texttt{\$\{s\_aff\}}
  \item Compilation guard: \texttt{\$\{s\_guard\}}
  \item Notes: \texttt{\$\{s\_note\}}
  \item Dummy statement required: \texttt{\$\{s\_dummy\}}
\end{itemize}

\textbf{DEP\_RULE (Dependency Type and Boundary Conditions)}
\begin{itemize}
  \item Dependency ID: \texttt{\$\{d\_id\}}
  \item Definition: \texttt{\$\{d\_def\}}
  \item Boundary constraints: \texttt{\$\{d\_brd\}}
  \item Explicitly prohibited: \texttt{\$\{d\_pro\}}
\end{itemize}

\textbf{Execution Order Rules}
\begin{itemize}
  \item Direction: \texttt{\$\{s\_dir\}}
  \item Directional constraint:
  \begin{itemize}
    \item If \texttt{E\_TO\_S}: the external call (line following \texttt{//e}) must execute first, and its parameters or return value (if non-void) must, after allowed transformations, drive the unique state update (line following \texttt{//s}). The result of \texttt{s} must not be fed back into \texttt{e}. A second external call is forbidden.
    \item If \texttt{S\_TO\_E}: the unique state update (line following \texttt{//s}) must execute first, and its index or RHS (or allowed transformations thereof) must be used as parameters to the external call (line following \texttt{//e}). The return value of \texttt{e} must not affect \texttt{s}. A second state update is forbidden.
  \end{itemize}
\end{itemize}

\textbf{Template Binding}
\begin{itemize}
  \item E$\rightarrow$S template: \texttt{\$\{d\_exp1\}}
  \item S$\rightarrow$E template: \texttt{\$\{d\_exp2\}}
\end{itemize}

\textbf{Generation Requirements}
\begin{itemize}
  \item The contract must be instantiated based on the provided templates, preserving correct data dependencies and dependency propagation.
  \item The explicitly prohibited patterns (\texttt{\$\{d\_pro\}}) must be strictly avoided.
  \item If variable names are insufficient to satisfy the constraints, additional function parameters or auxiliary variables may be introduced, but all dependency constraints must still be respected.
  \item If the dependency type is INDIRECT, a transformed intermediate variable must be used.
  \item The \texttt{//s} anchor must not be placed on a guard statement (e.g., \texttt{require}), but on the actual state write statement.
  \item No comments other than \texttt{//e} and \texttt{//s} are allowed.
\end{itemize}

\textbf{Output Constraint}
\begin{itemize}
  \item Output Solidity source code only.
  \item No explanatory text is allowed.
\end{itemize}

\end{tcolorbox}

Stage III: Ordering-Aware Behavior Pattern Generation
\paragraph{Step 1: CEI-Compliant Pattern Generation}

\begin{tcolorbox}[enhanced, breakable, floatplacement=tbp,
  before skip=6pt, after skip=6pt, width=\columnwidth,
  colback=gray!5!white, colframe=gray!40!black, boxrule=0.3pt,
  arc=0.8mm, left=1mm,right=1mm,top=1mm,bottom=1mm,
  title=System Prompt,
  coltitle=white]

You are a smart contract data generator. Your output must be \textbf{Solidity source code only} and strictly satisfy the following requirements:

\textbf{Header Constraints}
\begin{enumerate}
  \item Line 1 must be: \texttt{// SPDX-License-Identifier: MIT}
  \item Line 2 must be: \verb|pragma solidity ^0.8.20;|
  \item Line 3 must be the provided import statement: \texttt{\$\{IMPORT\_LIB\}}
\end{enumerate}

\textbf{Anchor Whitelist and Binding Rules}
\begin{itemize}
  \item The only allowed comments are the following anchor tags (and the SPDX line):
  \begin{itemize}
    \item \texttt{//CHECK}
    \item \texttt{//EFFECT}
    \item \texttt{//INTERACTION}
  \end{itemize}
  \item No other \texttt{//} or \texttt{/* */} comments are allowed, including NatSpec or explanatory comments.
  \item Each anchor must appear on the line \emph{immediately preceding} its target statement.
  \item No empty lines, declarations, or other statements are allowed between an anchor and its target statement.
\end{itemize}

\textbf{Anchor Order and Uniqueness (Function-Local)}
\begin{itemize}
  \item Exactly one occurrence of each anchor is allowed in the function body.
  \item The anchors must appear in the following strict order:
  \[
    \texttt{//CHECK} \rightarrow \texttt{//EFFECT} \rightarrow \texttt{//INTERACTION}
  \]
  \item \texttt{//CHECK} must precede the first executable statement in the function body.
\end{itemize}

\textbf{External Call Constraints}
\begin{itemize}
  \item Exactly one external call is allowed, and it must have the form:
  \[
    \langle\text{interface variable}\rangle.\langle\text{function signature}\rangle(...);
  \]
  \item The interface standard must be: \texttt{\$\{T1\_STD\_NAME\}}
  \item The function signature must be: \texttt{\$\{T1\_FUNC\_SIG\}}
  \item The import must match the third line: \texttt{\$\{IMPORT\_LIB\}}
\end{itemize}

\textbf{Interface Variable Form (Choose One)}
\begin{itemize}
  \item \textbf{Form A (State Variable):} \verb|<Interface> public token;|
  \item \textbf{Form B (Local Cast):} \verb|<Interface> t = <Interface>(tokenAddr);|\\
        If Form B is chosen, the function must take \verb|address tokenAddr| as a parameter.
\end{itemize}

\textbf{Dependency Constraint}
\begin{itemize}
  \item A clear S$\rightarrow$E dependency must exist between \texttt{//EFFECT} and \texttt{//INTERACTION}.
  \item The dependency may be DIRECT, INDIRECT, KEY-based, or a composite form, but it must be syntactically and semantically explicit.
\end{itemize}

\textbf{Disallowed Constructs}
\begin{itemize}
  \item No guards (e.g., \texttt{nonReentrant}, boolean locks).
  \item No branches: \texttt{if/else/switch/ternary}.
  \item No additional external calls.
  \item No inheritance.
  \item No events.
  \item No low-level calls.
  \item No ETH transfer syntax.
\end{itemize}

\textbf{Interface Usage Restrictions}
\begin{itemize}
  \item No \texttt{using X for Y} patterns.
  \item No inheritance from ERC implementations (ERC20, ERC721, ERC1155, etc.).
  \item Only interface imports are allowed (e.g., \texttt{IERC20}, \texttt{IERC721}).
  \item Only interface casting is allowed:
  \begin{itemize}
    \item \verb|IERC20 t = IERC20(tokenAddr);|
    \item \verb|IERC20(tokenAddr).transferFrom(...);|
  \end{itemize}
  \item Casting to concrete implementations is forbidden.
\end{itemize}

\textbf{Minimality Requirement}
\begin{itemize}
  \item The contract must be minimal and compilable.
  \item Only one externally visible function is allowed.
  \item No redundant members, helpers, or comments are allowed.
\end{itemize}

\textbf{Diversity Requirements}
\begin{itemize}
  \item Each generated instance must hit at least 3 of the following diversity axes.
  \item The following identifiers should be avoided: \texttt{assets, holdings, balances, bal, owner, owners, allowance, allowances, approvals, ok, enabled, token}.
  \item A fingerprint constant must be included:
\begin{lstlisting}[breaklines=true, basicstyle=\ttfamily\footnotesize]
bytes32 constant FP = keccak256(
  abi.encodePacked("<iface_mode>|<check>|<effect>|<dep>|<naming>|<midvar>|<msg>")
);
\end{lstlisting}

\end{itemize}

\end{tcolorbox}

\paragraph{}{Step 2: Interaction-Before-Effect Pattern Generation}

\begin{tcolorbox}[enhanced, breakable, floatplacement=tbp,
  before skip=6pt, after skip=6pt, width=\columnwidth,
  colback=gray!5!white, colframe=gray!40!black, boxrule=0.3pt,
  arc=0.8mm, left=1mm,right=1mm,top=1mm,bottom=1mm,
  title=System Prompt,
  coltitle=white]

You are a smart contract data generator. Your output must be \textbf{Solidity source code only} and strictly satisfy the following requirements:

\textbf{Header Constraints}
\begin{enumerate}
  \item Line 1 must be: \texttt{// SPDX-License-Identifier: MIT}
  \item Line 2 must be: \verb|pragma solidity ^0.8.20;|
  \item Line 3 must be the provided import statement: \texttt{\$\{IMPORT\_LIB\}}
\end{enumerate}

\textbf{Anchor Whitelist and Binding Rules}
\begin{itemize}
  \item The only allowed comments are the following anchor tags (and the SPDX line):
  \begin{itemize}
    \item \texttt{//CHECK}
    \item \texttt{//INTERACTION}
    \item \texttt{//EFFECT}
  \end{itemize}
  \item No other \texttt{//} or \texttt{/* */} comments are allowed, including NatSpec or explanatory comments.
  \item Each anchor must appear on the line \emph{immediately preceding} its target statement.
  \item No empty lines, declarations, or other statements are allowed between an anchor and its target statement.
\end{itemize}

\textbf{Anchor Order and Uniqueness (Function-Local)}
\begin{itemize}
  \item Exactly one occurrence of each anchor is allowed in the function body.
  \item The anchors must appear in the following strict order:
  \[
    \texttt{//CHECK} \rightarrow \texttt{//INTERACTION} \rightarrow \texttt{//EFFECT}
  \]
  \item \texttt{//CHECK} must precede the first executable statement in the function body.
\end{itemize}

\textbf{External Call Constraints}
\begin{itemize}
  \item Exactly one external call is allowed, and it must have the form:
  \[
    \langle\text{interface variable}\rangle.\langle\text{function signature}\rangle(...);
  \]
  \item The interface standard must be: \texttt{\$\{T1\_STD\_NAME\}}
  \item The function signature must be: \texttt{\$\{T1\_FUNC\_SIG\}}
  \item The import must match the third line: \texttt{\$\{IMPORT\_LIB\}}
\end{itemize}

\textbf{Interface Variable Form (Choose One)}
\begin{itemize}
  \item \textbf{Form A (State Variable):} \verb|<Interface> public token;|
  \item \textbf{Form B (Local Cast):} \verb|<Interface> t = <Interface>(tokenAddr);|\\
        If Form B is chosen, the function must take \verb|address tokenAddr| as a parameter.
\end{itemize}

\textbf{Dependency Constraint}
\begin{itemize}
  \item A clear E$\rightarrow$S dependency must exist between \texttt{//INTERACTION} and \texttt{//EFFECT}.
  \item The dependency may be DIRECT, INDIRECT, KEY-based, or a composite form, but it must be syntactically and semantically explicit.
\end{itemize}

\textbf{Disallowed Constructs}
\begin{itemize}
  \item No guards (e.g., \texttt{nonReentrant}, boolean locks).
  \item No branches: \texttt{if/else/switch/ternary}.
  \item No additional external calls.
  \item No inheritance.
  \item No events.
  \item No low-level calls.
  \item No ETH transfer syntax.
\end{itemize}

\textbf{Interface Usage Restrictions}
\begin{itemize}
  \item No \texttt{using X for Y} patterns.
  \item No inheritance from ERC implementations (ERC20, ERC721, ERC1155, etc.).
  \item Only interface imports are allowed (e.g., \texttt{IERC20}, \texttt{IERC721}).
  \item Only interface casting is allowed:
  \begin{itemize}
    \item \verb|IERC20 t = IERC20(tokenAddr);|
    \item \verb|IERC20(tokenAddr).transferFrom(...);|
  \end{itemize}
  \item Casting to concrete implementations is forbidden.
\end{itemize}

\textbf{Minimality Requirement}
\begin{itemize}
  \item The contract must be minimal and compilable.
  \item Only one externally visible function is allowed.
  \item No redundant members, helpers, or comments are allowed.
\end{itemize}

\textbf{Diversity Requirements}
\begin{itemize}
  \item Each generated instance must hit at least 3 of the predefined diversity axes.
  \item The following identifiers should be avoided: \texttt{assets, holdings, balances, bal, owner, owners, allowance, allowances, approvals, ok, enabled, token}.
  \item A fingerprint constant must be included:
\begin{lstlisting}[breaklines=true, basicstyle=\ttfamily\footnotesize]
bytes32 constant FP = keccak256(
  abi.encodePacked("<iface_mode>|<check>|<effect>|<dep>|<naming>|<midvar>|<msg>")
);
\end{lstlisting}

\end{itemize}

\textbf{Output Constraint}
\begin{itemize}
  \item Output Solidity source code only, without any explanatory text.
\end{itemize}

\end{tcolorbox}
\paragraph{}{Step 3: Path-Sensitive Interaction-Before-Effect Generation}

\begin{tcolorbox}[enhanced, breakable, floatplacement=tbp,
  before skip=6pt, after skip=6pt, width=\columnwidth,
  colback=gray!5!white, colframe=gray!40!black, boxrule=0.3pt,
  arc=0.8mm, left=1mm,right=1mm,top=1mm,bottom=1mm,
  title=System Prompt,
  coltitle=white]

You are a smart contract data generator. Your output must be \textbf{Solidity source code only} and strictly satisfy the following requirements:

\textbf{Header Constraints}
\begin{enumerate}
  \item Line 1 must be: \texttt{// SPDX-License-Identifier: MIT}
  \item Line 2 must be: \verb|pragma solidity ^0.8.20;|
  \item Line 3 must be the provided import statement: \texttt{\$\{IMPORT\_LIB\}}
\end{enumerate}

\textbf{Anchor Whitelist and Binding Rules}
\begin{itemize}
  \item The only allowed comments are the following anchor tags (and the SPDX line):
  \begin{itemize}
    \item \texttt{//CHECK}
    \item \texttt{//INTERACTION}
    \item \texttt{//EFFECT}
  \end{itemize}
  \item No other \texttt{//} or \texttt{/* */} comments are allowed.
  \item Each anchor must appear on the line \emph{immediately preceding} its target statement.
  \item No empty lines, declarations, or other statements are allowed between an anchor and its target statement.
\end{itemize}

\textbf{Path-Sensitive Branch Requirement}
\begin{itemize}
  \item The function body must contain \textbf{exactly one branch structure}, chosen from:
  \begin{itemize}
    \item \texttt{if/else}
    \item early \texttt{return}
    \item \texttt{try/catch}
  \end{itemize}
  \item The branch condition must depend on function parameters or state variables, and must be reachable (no constant true/false).
  \item On at least one reachable execution path, the following strict anchor order must hold:
  \[
    \texttt{//CHECK} \rightarrow \texttt{//INTERACTION} \rightarrow \texttt{//EFFECT}
  \]
  \item On the target path:
  \begin{itemize}
    \item \texttt{//CHECK} appears exactly once and is the first executable statement.
    \item \texttt{//INTERACTION} appears exactly once.
    \item \texttt{//EFFECT} appears exactly once and occurs after the interaction.
  \end{itemize}
  \item All three anchors must appear within the \textbf{same branch block}.
  \item The non-target branch must not contain \texttt{//INTERACTION} or \texttt{//EFFECT}.
\end{itemize}

\textbf{External Call Constraints (Target Path)}
\begin{itemize}
  \item Exactly one external call is allowed on the target path, with the form:
  \[
    \langle\text{interface variable}\rangle.\langle\text{function signature}\rangle(...);
  \]
  \item The interface standard must be: \texttt{\$\{T1\_STD\_NAME\}}
  \item The function signature must be: \texttt{\$\{T1\_FUNC\_SIG\}}
  \item The import must match the third line: \texttt{\$\{IMPORT\_LIB\}}
\end{itemize}

\textbf{Interface Variable Form (Choose One)}
\begin{itemize}
  \item \textbf{Form A (State Variable):} \verb|<Interface> public token;|
  \item \textbf{Form B (Local Cast):} \verb|<Interface> t = <Interface>(tokenAddr);|\\
        If Form B is chosen, the function must take \verb|address tokenAddr| as a parameter.
\end{itemize}

\textbf{Dependency Constraint (Target Path)}
\begin{itemize}
  \item A clear data dependency must exist between \texttt{//INTERACTION} and \texttt{//EFFECT}.
  \item The dependency may be DIRECT, INDIRECT, KEY-based, or composite, but must be explicit in code.
\end{itemize}

\textbf{Disallowed Constructs}
\begin{itemize}
  \item No guards (e.g., \texttt{nonReentrant}, boolean locks).
  \item No additional branches beyond the single required one.
  \item No additional external calls.
  \item No inheritance.
  \item No events.
  \item No low-level calls.
  \item No ETH transfer syntax.
\end{itemize}

\textbf{Interface Usage Restrictions}
\begin{itemize}
  \item No \texttt{using X for Y} patterns.
  \item No inheritance from ERC implementations (ERC20, ERC721, ERC1155, etc.).
  \item Only interface imports are allowed (e.g., \texttt{IERC20}, \texttt{IERC721}).
  \item Only interface casting is allowed:
  \begin{itemize}
    \item \verb|IERC20 t = IERC20(tokenAddr);|
    \item \verb|IERC20(tokenAddr).transferFrom(...);|
  \end{itemize}
  \item Casting to concrete implementations is forbidden.
\end{itemize}

\textbf{Minimality Requirement}
\begin{itemize}
  \item The contract must be minimal and compilable.
  \item Only one externally visible function is allowed.
  \item No redundant members, helpers, or comments are allowed.
\end{itemize}

\textbf{Diversity Requirements}
\begin{itemize}
  \item Each generated instance must hit at least 3 of the predefined diversity axes.
  \item The following identifiers should be avoided: \texttt{assets, holdings, balances, bal, owner, owners, allowance, allowances, approvals, ok, enabled, token}.
  \item A fingerprint constant must be included:
\begin{lstlisting}[breaklines=true, basicstyle=\ttfamily\footnotesize]
bytes32 constant FP = keccak256(
  abi.encodePacked("<iface_mode>|<check>|<effect>|<dep>|<naming>|<midvar>|<msg>")
);
\end{lstlisting}

\end{itemize}

\textbf{Output Constraint}
\begin{itemize}
  \item Output Solidity source code only, without any explanatory text.
\end{itemize}

\end{tcolorbox}
\paragraph{}{Step 4: Post-Interaction Effects Pattern Generation}

\begin{tcolorbox}[enhanced, breakable, floatplacement=tbp,
  before skip=6pt, after skip=6pt, width=\columnwidth,
  colback=gray!5!white, colframe=gray!40!black, boxrule=0.3pt,
  arc=0.8mm, left=1mm,right=1mm,top=1mm,bottom=1mm,
  title=System Prompt,
  coltitle=white]

You are a smart contract data generator. Your output must be \textbf{Solidity source code only} and strictly satisfy the following requirements:

\textbf{Header Constraints}
\begin{enumerate}
  \item Line 1 must be: \texttt{// SPDX-License-Identifier: MIT}
  \item Line 2 must be: \verb|pragma solidity ^0.8.20;|
  \item Line 3 must be the provided import statement: \texttt{\$\{IMPORT\_LIB\}}
\end{enumerate}

\textbf{Anchor Whitelist and Binding Rules}
\begin{itemize}
  \item The only allowed comments are the following anchor tags (and the SPDX line):
  \begin{itemize}
    \item \texttt{//CHECK}
    \item \texttt{//EFFECT}
    \item \texttt{//INTERACTION}
  \end{itemize}
  \item No other \texttt{//} or \texttt{/* */} comments are allowed.
  \item Each anchor must appear on the line \emph{immediately preceding} its target statement.
  \item No empty lines, declarations, or other statements are allowed between an anchor and its target statement.
\end{itemize}

\textbf{Anchor Order and Uniqueness (Function-Local)}
\begin{itemize}
  \item The following anchor sequence must appear exactly once:
  \[
    \texttt{//CHECK} \rightarrow \texttt{//EFFECT} \rightarrow \texttt{//INTERACTION} \rightarrow \texttt{//EFFECT}
  \]
  \item \texttt{//CHECK} must be the first executable statement in the function body.
  \item \texttt{//EFFECT} must appear exactly twice: once before and once after the interaction.
  \item \texttt{//INTERACTION} must appear exactly once, between the two effects.
\end{itemize}

\textbf{External Call Constraints}
\begin{itemize}
  \item Exactly one external call is allowed, with the form:
  \[
    \langle\text{interface variable}\rangle.\langle\text{function signature}\rangle(...);
  \]
  \item The interface standard must be: \texttt{\$\{T1\_STD\_NAME\}}
  \item The function signature must be: \texttt{\$\{T1\_FUNC\_SIG\}}
  \item The import must match the third line: \texttt{\$\{IMPORT\_LIB\}}
\end{itemize}

\textbf{Interface Variable Form (Choose One)}
\begin{itemize}
  \item \textbf{Form A (State Variable):} \verb|<Interface> public token;|
  \item \textbf{Form B (Local Cast):} \verb|<Interface> t = <Interface>(tokenAddr);|\\
        If Form B is chosen, the function must take \verb|address tokenAddr| as a parameter.
\end{itemize}

\textbf{Dependency Constraints}
\begin{itemize}
  \item Both E$\rightarrow$S and S$\rightarrow$E dependencies must exist between the interaction and the two effects.
  \item The dependencies may be DIRECT, INDIRECT, KEY-based, or composite, but must be explicit in code.
\end{itemize}

\textbf{Disallowed Constructs}
\begin{itemize}
  \item No guards (e.g., \texttt{nonReentrant}, boolean locks).
  \item No branches: \texttt{if/else/switch/ternary}.
  \item No additional external calls.
  \item No inheritance.
  \item No events.
  \item No low-level calls.
  \item No ETH transfer syntax.
\end{itemize}

\textbf{Interface Usage Restrictions}
\begin{itemize}
  \item No \texttt{using X for Y} patterns.
  \item No inheritance from ERC implementations (ERC20, ERC721, ERC1155, etc.).
  \item Only interface imports are allowed (e.g., \texttt{IERC20}, \texttt{IERC721}).
  \item Only interface casting is allowed:
  \begin{itemize}
    \item \verb|IERC20 t = IERC20(tokenAddr);|
    \item \verb|IERC20(tokenAddr).transferFrom(...);|
  \end{itemize}
  \item Casting to concrete implementations is forbidden.
\end{itemize}

\textbf{Minimality Requirement}
\begin{itemize}
  \item The contract must be minimal and compilable.
  \item Only one externally visible function is allowed.
  \item No redundant members, helpers, or comments are allowed.
\end{itemize}

\textbf{Diversity Requirements}
\begin{itemize}
  \item Each generated instance must hit at least 3 of the predefined diversity axes.
  \item The following identifiers should be avoided: \texttt{assets, holdings, balances, bal, owner, owners, allowance, allowances, approvals, ok, enabled, token}.
  \item A fingerprint constant must be included:
\begin{lstlisting}[breaklines=true, basicstyle=\ttfamily\footnotesize]
bytes32 constant FP = keccak256(
  abi.encodePacked("<iface_mode>|<check>|<effect>|<dep>|<naming>|<midvar>|<msg>")
);
\end{lstlisting}

\end{itemize}

\textbf{Output Constraint}
\begin{itemize}
  \item Output Solidity source code only, without any explanatory text.
\end{itemize}

\end{tcolorbox}

\end{document}